\documentclass[twocolumn,showpacs,nofootinbib,preprintnumbers,amsmath,amssymb,showkeys,superscriptaddress]{revtex4}

\usepackage{epsfig}
\usepackage{dsfont}
\usepackage{color}
\usepackage{graphicx}
\usepackage{dcolumn}
\usepackage{bm}

\def\slashii#1{\setbox0=\hbox{$#1$}             
   \dimen0=\wd0                                 
   \setbox1=\hbox{\sl/} \dimen1=\wd1            
   \ifdim\dimen0>\dimen1                        
      \rlap{\hbox to \dimen0{\hfil\sl/\hfil}}   
      #1                                        
   \else                                        
      \rlap{\hbox to \dimen1{\hfil$#1$\hfil}}   
      \hbox{\sl/}                               
   \fi}                                         %
\def\slashiii#1{\setbox0=\hbox{$#1$}#1\hskip-\wd0\hbox to\wd0{\hss\sl/\/\hss}}
\newcommand{\eq}[1]{\begin{align} #1 \end{align}}
\newcommand{\beq}{\begin{equation}}
\newcommand{\eeq}{\end{equation}}
\newcommand{\bea}{\begin{eqnarray}}
\newcommand{\eea}{\end{eqnarray}}
\newcommand{\nn}{\nonumber \\}

\newcommand\eqn[1]{(\ref{#1})}      
\newcommand\Eqn[1]{Eq.~(\ref{#1})}  
\newcommand\Fig[1]{Fig.~\ref{#1}}  
\newcommand\Sec[1]{Sec.~\ref{#1}}  

\begin{document}


\title{How nonperturbative is the infrared regime of Landau gauge\\ Yang-Mills correlators?}

\author{U. Reinosa}%
\affiliation{%
Centre de Physique Th\'eorique, Ecole Polytechnique, CNRS, Universit\'e Paris-Saclay, F-91128 Palaiseau, France
}%
\author{J. Serreau}%
\affiliation{%
 Astro-Particule et Cosmologie (APC), CNRS UMR 7164, Universit\'e Paris 7 - Denis Diderot,\\ 10, rue Alice Domon et L\'eonie Duquet, 75205 Paris Cedex 13, France
}%
\author{M. Tissier}
\affiliation{LPTMC, Laboratoire de Physique Th\'eorique de la Mati\`ere Condens\'ee, CNRS UMR 7600, Universit\'e Pierre et Marie Curie, \\ boite 121, 4 pl. Jussieu, 75252 Paris Cedex 05, France
}
\author{N. Wschebor}%
\affiliation{%
 Instituto de F\'{\i}sica, Facultad de Ingenier\'{\i}a, Universidad de la Rep\'ublica,\\ J.H.y Reissig 565, 11000 Montevideo, Uruguay
}%

\date{\today}

\begin{abstract}
{We study the Landau gauge correlators of Yang-Mills fields for infrared Euclidean momenta in the context of a massive extension of the Faddeev-Popov Lagrangian which, we argue, underlies a variety of continuum approaches. Standard (perturbative) renormalization group techniques with a specific, infrared-safe renormalization scheme produce so-called decoupling and scaling solutions for the ghost and gluon propagators, which correspond to nontrivial infrared fixed points. The decoupling fixed point is infrared stable and weakly coupled, while the scaling fixed point is unstable and generically strongly coupled except for low dimensions $d\to2$. Under the assumption that such a scaling fixed point exists beyond one-loop order, we find that the corresponding ghost and gluon scaling exponents are, respectively, $2\alpha_F=2-d$ and $2\alpha_G=d$ at all orders of perturbation theory in the present renormalization scheme. We discuss the relation between the ghost wave function renormalization, the gluon screening mass, the scale of spectral positivity violation, and  the gluon mass parameter. We also show that this scaling solution does not realize the standard Becchi-Rouet-Stora-Tyutin symmetry of the Faddeev-Popov Lagrangian. Finally, we discuss our findings in relation to the results of nonperturbative continuum methods. }
 \end{abstract}

\maketitle


\section{Introduction}
\label{sec:intro}

Understanding the behavior of the correlation functions of Yang-Mills (YM) fields for infrared momenta is of key importance, in particular, for continuum approaches to the dynamics of strong interactions. In the past two decades, intense efforts have been devoted to compute the Landau gauge\footnote{The Coulomb gauge has also been the subject of many studies \cite{Gribov77,Cucchieri:2000gu,Langfeld:2004qs,Feuchter:2004mk}. The infrared behavior of correlators is quite different from that in the Landau gauge, and we shall not discuss it here.} Yang-Mills ghost and gluon correlators for infrared Euclidean momenta \cite{Alkofer00,Boucaud:2011ug}. Lattice studies \cite{Bonnet:2000kw,Cucchieri_08b,Bogolubsky09,Bornyakov09,Iritani:2009mp,Maas:2011se,Oliveira:2012eh} have unambiguously demonstrated that, in $d=3$ and $d=4$ dimensions, the gluon propagator saturates to a finite value at vanishing momentum, corresponding to a nonzero screening mass, and shows a violation of spectral positivity, which indicates that the corresponding massivelike excitation does not correspond to an asymptotic state, as expected from confinement. At the same time, the ghost dressing function is finite for all momenta, which realizes a particular case of the class of so-called decoupling solutions. Noticeably, the gauge coupling, extracted from the gluon-ghost-antighost vertex, stays finite for all momenta and even vanishes in the deep infrared \cite{Bogolubsky09,Boucaud:2011ug}. The situation is different in $d=2$, where the regime of infrared momenta is characterized by a scaling-type solution, with a vanishing gluon propagator (i.e., an infinite screening mass), a power law divergent ghost dressing function, and a finite, nonzero ghost-gluon coupling \cite{Maas:2007uv,Cucchieri:2011um,Cucchieri:2011ig}. Both the decoupling and the scaling solutions are clearly at odds with standard perturbation theory, based on the Faddeev-Popov (FP) quantization procedure, which is plagued by a Landau pole for infrared momenta, where the coupling grows without bound. 

A wide variety of continuum approaches has been developed to tackle this issue in the Landau gauge. These include nonperturbative approximation schemes based on truncations of the hierarchy of Dyson-Schwinger (DSE) \cite{vonSmekal97,Atkinson:1998zc,Zwanziger:2001kw,Lerche:2002ep,Fischer:2002hna,Maas:2004se,Boucaud06,Huber:2007kc,Aguilar07,Aguilar08,Boucaud:2008ji,Dall'Olio:2012zw,Huber:2012zj,Huber:2016tvc} or functional renormalization group (FRG) \cite{Ellwanger96,Pawlowski:2003hq,Fischer:2004uk,Fischer:2006vf,Fischer08,Cyrol:2016tym} equations as well as variational methods in the Hamiltonian formalism (HF) \cite{Schleifenbaum:2006bq,Quandt:2013wna,Quandt:2015aaa}. Other approaches are based on modified quantization schemes, which deal with the issue of Gribov ambiguities (namely the fact that the Landau gauge condition fixes the gauge only up to a discrete set of equivalent copies), such as the (refined) Gribov-Zwanziger approach \cite{Gribov77,Zwanziger89,Vandersickel:2012tz,Dudal:2008sp} or the massive extension of the FP Lagrangian \cite{Tissier:2010ts,Tissier:2011ey} based on the quantization procedure of Ref.~\cite{Serreau:2012cg}. These methods are all able to produce various decoupling and scaling solutions.

Two related points are worth emphasizing here. First, the nonperturbative continuum approaches mentioned above have to adjust one extra parameter on top of the gauge coupling in order to reproduce the lattice results for the ghost and gluon correlators. This is, for instance, a boundary condition for the ghost dressing functions in DSE studies \cite{Zwanziger:2001kw,Lerche:2002ep,Fischer08} or an ultraviolet (UV) gluon mass parameter in FRG works \cite{Fischer:2004uk,Fischer08,Cyrol:2016tym}. Second, these approaches are not directly based on the sole FP Lagrangian. This is because it is not known, to date, how to keep track of the local Becchi-Rouet-Stora-Tyutin (BRST) symmetry of the latter beyond perturbation theory~\cite{Neuberger:1986vv}.\footnote{ We mention that a generalization of the BRST symmetry has been recently discovered in the context of the Gribov-Zwanziger approach \cite{Capri:2015ixa,Capri:2016gut}. However, this concerns a modified FP action.}

In fact, gauge-fixed lattice simulations are not based on the FP Lagrangian either because of the Gribov ambiguities. Various ways of dealing with Gribov copies have been considered \cite{Maas:2016frv}, the simplest one consisting in randomly picking up one copy on each gauge orbit, the so-called minimal Landau gauge. In any case, existing numerical algorithms are efficient only in the first Gribov region (where the FP operator is positive definite), a restriction which explicitly breaks the BRST symmetry of the FP Lagrangian.\footnote{The possibility that the FP construction for the Landau gauge is correct at a nonperturbative level even in presence of Gribov copies has been suggested \cite{Hirschfeld:1978yq,vonSmekal:2013cla,vonSmekal:2008en} but remains, for the moment, unproven.} 
 It is thus not surprising that the nonperturbative continuum approaches mentioned above have to adjust (at least) one extra parameter to describe actual lattice data. This is usually understood as an effective, {\it a posteriori} way to fix the residual gauge freedom, although we stress that, despite some attempts to put this idea on more solid grounds \cite{Maas:2008ri}, this remains to be firmly established. It is also not completely surprising that continuum approaches can reach a whole class of solutions that differ from lattice results, since they can explore a wider range of ``gauge-fixed'' (in the loose sense described above) Lagrangians. 

 In parallel to trying to reproduce lattice data in the minimal Landau gauge, one may also want to explore the possibility to produce results with the actual BRST--symmetric FP Lagrangian. To deal with the explicit breaking of the BRST symmetry induced by the regularization procedure, one should include, in the regularized Lagrangian defined at some UV scale, a whole set of relevant BRST breaking (couter)terms and adjust them so that  the BRST symmetry is recovered when the regulator is removed. We stress again that whether this procedure makes sense beyond perturbation theory is a nontrivial open question.
Moreover, there is the ``in principle" versus ``in practice" issue. In principle, the procedure described here imposes one to include all BRST--breaking terms allowed by power counting \cite{Becchi93,Ellwanger:1994iz,Lowenstein:1975pd,Meyers:2014iwa}. In practice, however, existing studies essentially include, for technical reasons, the minimal ingredient necessary to deal with quadratic divergences introduced by the BRST--breaking UV regulator \cite{Fischer:2004uk,Quandt:2013wna,Cyrol:2016tym,Meyers:2014iwa,Huber:2014tva}, which amounts to a gluon mass (counter)term.\footnote{ Some studies also include different couplings for the three-gluon, four-gluon, and ghost-gluon vertices \cite{Meyers:2014iwa,Cyrol:2016tym}, but these do not play an important role for the present discussion. In particular, the transition from decoupling to scaling solutions is essentially triggered by the gluon mass term.} The hope is then that there exists a unique value of the latter which exactly cancels the BRST breaking effect of the regulator. It has been conjectured that this corresponds to a scaling solution \cite{Lerche:2002ep,Fischer08}. We stress again that, although appealing, this scenario remains, at present, hypothetical. 

We see that, for all practical purposes, existing nonperturbative works are effectively based on a massive deformation of the FP Lagrangian, equivalent to the Landau limit of the Curci-Ferrari (CF) Lagrangian \cite{Curci:1976bt}, which has one more (dimensionful) parameter than the original YM theory. The question is, therefore, whether there exists a range of parameter space where the CF model actually provides a sensible realization of YM theory, possibly including a BRST symmetric solution.

Another line of reasoning, initiated in Refs.~\cite{Tissier:2010ts,Tissier:2011ey,Weber:2011nw}, is to consider the massive CF Lagrangian, not as an unwanted albeit necessary deformation of the theory that one has to eventually get rid of, but, instead, as an actual sensible starting point to study the infrared regime of the Landau gauge YM correlators. Here, the gluon mass term is seen as the minimal (local and renormalizable in $d\le4$ dimensions) extension of the FP Lagrangian which takes into account the effective BRST breaking due to the Gribov problem.\footnote{The general idea is that the BRST symmetry of the FP Lagrangian is likely not to be realized in a complete gauge fixing due to the Gribov problem. Assuming that the resulting gauge-fixed theory can at all be (at least effectively) formulated as a local Lagrangian, this will result in corresponding deformations of the FP Lagrangian. The simplest modification which preserves the well-tested UV behavior of the FP Lagrangian is a gluon mass term.} An explicit realization of this model in the context of a gauge fixing procedure, which consistently deals with the Gribov ambiguities, has been discussed in Ref.~\cite{Serreau:2012cg}. 

It has been shown that, first, the lattice results for the ghost and gluon propagators in the Landau gauge can be accurately described by a simple one-loop calculation in the massive model \cite{Tissier:2010ts} and, second, that the latter possesses infrared-safe renormalization schemes, with no Landau pole, which allow for renormalization group (RG) improved perturbative studies of the infrared regime \cite{Tissier:2011ey}. In this context, the lattice results correspond to an infrared-safe trajectory where the relevant expansion parameter remains moderate along the flow \cite{Boucaud:2011ug}. This perturbative approach has been extended to the calculation of two- and three-point YM and QCD correlators \cite{Pelaez:2013cpa} and to nonzero temperature  applications \cite{Reinosa:2013twa,Reinosa:2014ooa,Reinosa:2015oua,Reinosa:2016iml}. Similar ideas have also been implemented in Refs.~\cite{Siringo:2015wtx,Machado:2016cij,Weber:2016biv}. This has the great advantage to rely on standard (and often simple) perturbative calculations, which can be systematically improved. Although the philosophical status of the gluon mass here is very different from the one mentioned above in the context of nonperturbative approaches, it is of great interest to study the possibility of scaling solutions and the various related questions mentioned above within this perturbative approach. This is the purpose of the present work.

We study in detail the parameter space of the massive theory by means of the perturbative infrared-safe RG approach. At one-loop order, the RG flow exhibits a rich structure with different phases, corresponding to either Landau pole or infrared-safe trajectories, separated by a transition line (separatrix), which relates the ultraviolet Gaussian fixed point to a nontrivial infrared fixed point \cite{Serreau:2012cg}. Trajectories in the infrared-safe phase correspond to the continuous family of decoupling solutions, while the trajectory corresponding to the separatrix yields a scaling solution. An interesting simplification of the infrared-safe renormalization scheme used here is that the ghost and gluon propagators assume particularly simple forms in terms of the running coupling and mass parameters, which allow for discussing various features of the solutions in a simple way. Moreover, thanks to dimensional regularization, we can explicitly keep track of the deformed BRST symmetry of the massive model. In particular, this forbids quadratic divergences and guarantees that the gluon mass is multiplicatively renormalized. Moreover, the corresponding modified Slavnov-Taylor (ST) identities pose constraints on the possible infrared (perturbative) solutions.

We establish various properties, valid at all orders of perturbation theory, under the only assumption of the existence of an infrared scaling fixed point away from the (massless) FP limit. For instance, we determine the exact values of the exponents describing the scaling behavior of the ghost and gluon propagators on the separatrix, valid in dimensions $2\le d\le4$. We explicitly check that the assumption of a nontrivial infrared fixed point is satisfied at one-loop order in this range. Our scaling exponents agree with some results from DSE studies \cite{Zwanziger:2001kw,Lerche:2002ep}, although the latter---as well as other nonperturbative approaches---typically also find other possible exponents. 

We also take advantage of the simplicity of our approach to discuss various questions raised in previous studies.\footnote{ As a word of caution, we emphasize that the present analysis leaves open the possibility of genuine nonperturbative solutions, not attainable by perturbative means. Clearly, our analysis and results do not apply to such cases.} In particular, we  compute explicitly at one-loop order the relation between the ghost dressing function at vanishing momentum and the gluon mass parameter. These are related to the control parameters of DSE and FRG/HF studies, respectively. We also investigate the dependence of the gluon screening mass squared (defined as the inverse correlator at vanishing momentum) and of the scale of spectral positivity violation in the gluon sector as functions of the gluon mass parameter. It has been proposed in Ref.~\cite{Cyrol:2016tym}  that these can be used to distinguish between two ``phases'', called ``confining'' and ``Higgs--like'' in this reference. Here, we find no sign of a sharp transition between qualitatively distinct phases, but rather a smooth crossover between quantitatively different regimes. 

Finally, we discuss the scaling solution in relation with the issue of the possible BRST symmetry restoration mentioned earlier. The modified ST identities of the massive model impose that the longitudinal component of the gluon two-point vertex function is proportional to the ghost dressing function and, thus, does not vanish (as would be required by the BRST symmetry), whatever the (nonzero) value of the gluon mass parameter. In particular, in our RG scheme, the longitudinal and transverse gluon screening masses are always proportional to each other and are thus both infinite for the scaling solution. More generally, we show that, for any renormalization scheme compatible with the modified ST identities, demanding an approximate restoration of the BRST symmetry (in the sense that the longitudinal screening mass be negligible as compared to the transverse one) in the infrared regime strongly constrains the exponents of a possible (perturbative) scaling solution.

The paper is organized as follows. We briefly review, in \Sec{sec:massive}, the massive model in the infrared-safe (perturbative) renormalization scheme and describe the general structure of the RG flow in $d=4$ at one-loop order. We recall that, in the ultraviolet limit, the running mass is strongly suppressed so that we recover the standard FP theory. In Sec.~\ref{sec:IRphase}, we analyze the infrared-safe phase in detail. Typical trajectories head towards a weakly coupled attractive fixed point in the infrared, corresponding to the decoupling solutions for the ghost and gluon propagators~\cite{Tissier:2011ey}. Instead, the critical trajectory corresponding to the separatrix ends at a nontrivial infrared fixed point, which corresponds to a scaling solution of the Gribov type \cite{Gribov77}. Although the scaling fixed point is at strong coupling, we obtain exact values for the scaling exponents, valid at all orders of perturbation theory. This generalizes to arbitrary dimensions $2< d \le4$. We analyze in detail the case $d\to2$, where both the decoupling and the scaling fixed points are weakly coupled. Section \ref{sec:discussion} presents a discussion of the various properties of the infrared-safe solutions, in particular, concerning the ghost dressing function at zero momentum, the gluon screening mass, the scale of spectral positivity violation, and the question of BRST symmetry restoration. Finally, we discuss,  in \Sec{sec:comparison}, the results of our RG analysis concerning the scaling solution in relation with other (nonperturbative) continuum approaches. We conclude in \Sec{sec:concl}. For completeness, we recall how the present infrared-safe RG approach compares with the lattice data for the SU($3$) theory in $d=4$ in Appendix~\ref{appsec:lattice}. Finally, Appendix~\ref{app:toy} presents an illustrative toy DSE with perturbative and nonperturbative solutions.

\section{The massive Landau gauge}
\label{sec:massive}

As explained above, we consider a massive deformation of the standard FP Lagrangian in the Landau gauge. The latter is a particular case of the CF Lagrangian \cite{Curci:1976bt}. This model possesses a nontrivial phase structure in parameter space with, in particular, infrared-safe renormalization group trajectories~\cite{Tissier:2011ey,Serreau:2012cg}. Here, we briefly review the corresponding renormalization scheme and its actual implementation at one-loop order, focusing on what is relevant for the present analysis. The reader is referred to Ref.~\cite{Tissier:2011ey} for further details.

\subsection{Generalities}

We consider the Euclidean action $S=\int d^dx{\cal L}$ in $d$ dimensions, with
\eq{\label{eq:model}{\cal L}=\frac{1}{4}F_{\mu\nu}^aF_{\mu\nu}^a+\frac{m_B^2}{2}A_\mu^aA_\mu^a+ih^a\partial_\mu A_\mu^a+\partial_\mu\bar c^a(D_\mu c)^a,
}
where $A_\mu^a$ is the gauge field, $c^a$ and $\bar c^a$ a pair of ghost and antighost fields, and $h^a$ is a Nakanishi-Lautrup field, whose equation of motion enforces the Landau gauge condition $\partial_\mu A_\mu^a=0$. The covariant derivative $(D_\mu c)^a=\partial_\mu c^a+g_B f^{abc}A_\mu^b c^c$ and the
field strength tensor $F_{\mu\nu}^a=\partial_\mu A_\nu^a-\partial_\nu A_\mu^a+g_Bf^{abc}A_\mu^bA_\nu^c$ are expressed in terms of the bare coupling
constant $g_B$ and the bare mass parameter $m_B$. The latin indices correspond to the adjoint representation of the SU($N$) gauge group.

The model \eqn{eq:model} is multiplicatively renormalizable and possesses various symmetries which reduce the number of independent renormalization factors to two \cite{Curci:1976bt,Gracey:2002yt,Dudal:2002pq}.
We introduce renormalized fields and parameters as $A=\sqrt{Z_A}A_R$, $c=\sqrt{Z_c}c_R$, $\bar c=\sqrt{Z_c}\bar c_R$, $g_B=Z_g g$, and $m_B^2=Z_{m^2}m^2$. The renormalized ghost and gluon propagators are written as
\eq{
G^{ab}(p)&=\delta^{ab}\frac{F(p)}{p^2}, \\
G_{\mu\nu}^{ab}(p)&=\delta^{ab} P^\perp_{\mu\nu}(p) G(p),
}
with $P^\perp_{\mu\nu}(p)=\delta_{\mu\nu}-p_\mu p_\nu/p^2$. 
The function $F(p)$ is known as the ghost dressing function and we shall refer to $G(p)$ as the gluon propagator for simplicity. 

\subsection{The infrared-safe renormalization scheme}

Following Ref.~\cite{Tissier:2011ey}, we choose the following renormalization conditions for the two-point functions:
\eq{
\label{eq:renpres1}
 F(p=\mu)=1\,,\quad G^{-1}(p=\mu)=m^2+\mu^2,
}
and we further fix the values of the following finite combinations of renormalization factors as\footnote{That these combinations of renormalization factors are finite are consequences of nonrenormalization theorems which, themselves, follow from the modified ST identities of the massive model \cite{Taylor:1971ff,Curci:1976bt,Gracey:2002yt,Dudal:2002pq,Wschebor:2007vh,Tissier:2011ey}.}
\eq{
\label{eq:renpres2}
 Z_g \sqrt{Z_A} Z_c =1\,,\quad Z_{m^2} Z_A Z_c =1.
}
Renormalization group flows for the running parameters are obtained from these relations in a standard way. We introduce the coupling $\lambda=g^2N/(16\pi^2)$ and define the beta functions
\eq{\label{eq:betafunc}
\beta_{m^2}=\frac{d m^2}{d\ln\mu}\,,\quad
\beta_\lambda=\frac{d \lambda}{d\ln\mu},
}
and
\eq{
\gamma_A=\frac{d\log Z_A}{d\ln\mu}\,,\quad
\gamma_c=\frac{d\log Z_c}{d\ln\mu},
}
where the derivatives are taken at fixed bare parameters. With the renormalization prescriptions \eqn{eq:renpres2}, we have the relations
\eq{
\label{m2IRscheme}
\beta_{m^2}=m^2(\gamma_A+\gamma_c)\,,\quad\beta_\lambda=\lambda(\gamma_A+2\gamma_c).
}
It is important to remark that the functions $\gamma_A$ and $\gamma_c$ are finite in the FP theory so that the flow of the mass term vanishes in the limit of vanishing mass. This is because the BRST symmetry prevents the
appearance of a gluon mass. In particular, this guarantees that the UV flow of the mass is only logarithmic, such that the dimensionless ratio $m^2/\mu^2$ vanishes and one indeed recovers the correct UV behavior~\cite{Tissier:2011ey}; see also below. 

For later purposes, it is also useful to introduce the dimensionless quantity $\tilde m^2=m^2/\mu^2$, whose beta function depends only on
the renormalized parameters $\lambda$ and $\tilde m^2$ and reads
\eq{\label{eq:betatildem2}
\beta_{\tilde m^2}=\tilde m^2(-2+\gamma_A+\gamma_c).
}
Using the renormalization prescriptions \eqn{eq:renpres1} and choosing the renormalization scale $\mu=p$, one obtains the following expressions of the propagators\footnote{Specifically, making explicit the dependence on the RG scale and running parameters, we have, e.g., for the ghost dressing function, $F(p)\equiv F(p;\mu_0,m^2_0,\lambda_0)=z_c(\mu,\mu_0)F(p;\mu,m^2(\mu),\lambda(\mu))$, where the second equality uses the Callan-Symanzik equation, with $z_c(\mu,\mu_0)=\exp\int_{\mu_0}^\mu d\mu'\gamma_c(\mu')/\mu'$. Using $\mu=p$ and the renormalization conditions \eqn{eq:renpres1} yields $F(p)=z_c(p,\mu_0)$. The relations \eqn{m2IRscheme} then imply the result \eqn{eq:ghostprop}. The gluon propagator \eqn{eq:glueprop} is obtained along similar lines.}, in terms of the running parameters $m^2(\mu)$ and $\lambda(\mu)$:
\eq{\label{eq:ghostprop}
F(p)&=\frac{m^2_0}{\lambda_0}\frac{\lambda(p)}{m^2(p)}\,,\\
\label{eq:glueprop}
G(p)&=\frac{\lambda_0}{m^4_0}\frac{m^4(p)}{\lambda(p)}\frac{1}{p^2+m^2(p)},
}
where $m_0^2=m^2(\mu_0)$ and $\lambda_0=\lambda(\mu_0)$.

\begin{figure}[t!]
  \centering
  \includegraphics[width=.9\linewidth]{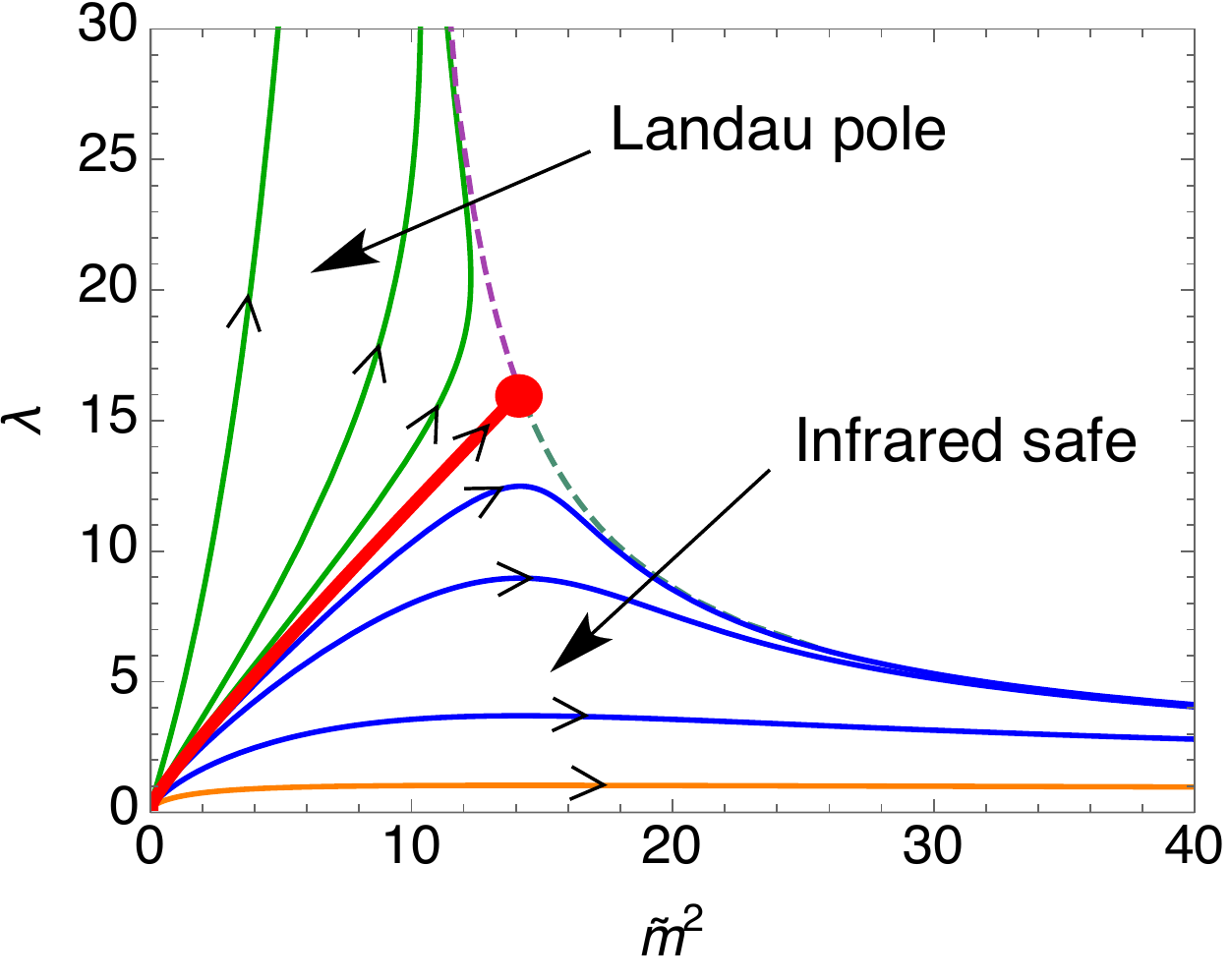}
  \caption{One-loop phase diagram and RG flow trajectories in the plane ($\tilde m^2, \lambda$) in $d=4$. The arrows indicate the flow towards the infrared. Trajectories which connect to the  ultraviolet Gaussian fixed point $(0,0)$ are separated in two classes: those which end at a Landau pole (green) and those which are infrared safe (blue), corresponding to decoupling solutions for the propagators. These are separated by a critical trajectory (red) which relates the Gaussian fixed point to a nontrivial infrared fixed point (red dot) at finite, nonzero values of $\tilde m^2$ and $\lambda$ and corresponds to a scaling solution for the correlators. We also show (orange, lower curve) the trajectory which describes lattice results for the SU($3$) theory (see Appendix~\ref{appsec:lattice}).}
  \label{fig_flow}
\end{figure}

It is important to realize that the expressions \eqn{eq:ghostprop} and \eqn{eq:glueprop} for the propagators are valid at all orders in pertubation theory. However, we cannot exclude that, for a given set $Z_A$, $Z_c$, $m_B$ and $g_B$ fixing the theory and the normalization of the fields, in addition to a perturbative solution that obeys all the renormalization conditions of the IR safe scheme, there exist genuine nonperturbative solutions such that some of the renormalization conditions, and in turn \eqn{eq:ghostprop} and \eqn{eq:glueprop}, are not obeyed. We illustrate this situation using a toy example in Appendix~\ref{app:toy}. The discussion of those solutions, if they exist, is beyond the scope of the present work, and we shall restrict to the study of the perturbative solutions. In particular, any result that follows from \eqn{eq:ghostprop} and \eqn{eq:glueprop} should be understood for the perturbative solution, even if it applies to all orders. In the following, we shall make statements about these perturbative solutions that are valid at all loop orders and others that will explicitly rely on a given approximation. We shall use the one-loop flow functions to illustrate our statements, keeping in mind the range of validity of this approximation in each case. 
The one-loop expressions of the gamma functions in the present infrared-safe renormalization scheme are \cite{Tissier:2011ey}
\eq{\label{eq:gammac}
\gamma_c=-\frac{\lambda}{2t^2} \Big[2t^2+2t -t^3 \ln t+(t+1)^2 (t-2) \ln (t+1)\Big],
}
where we denote $t= 1/\tilde m^2$, and
\eq{
\label{eq:gammaA}
\gamma_A&=\frac{\lambda}{6t^3}  \Bigg[-17 t^3+74t^2-12t+t^5 \ln t\nn
&-(t-2)^2 (2 t-3)
  (t+1)^2 \ln (t+1) \nn
&-t^{3\over2} \sqrt{t+4} \left(t^3-9
   t^2+20 t-36\right)\ln\!\left(\frac{\sqrt{t+4}-\sqrt{t}}{\sqrt{t+4}+\sqrt{t}}\right)\!\!\Bigg]\!.
}

\begin{figure}[t!]
  \centering
  \includegraphics[width=.9\linewidth]{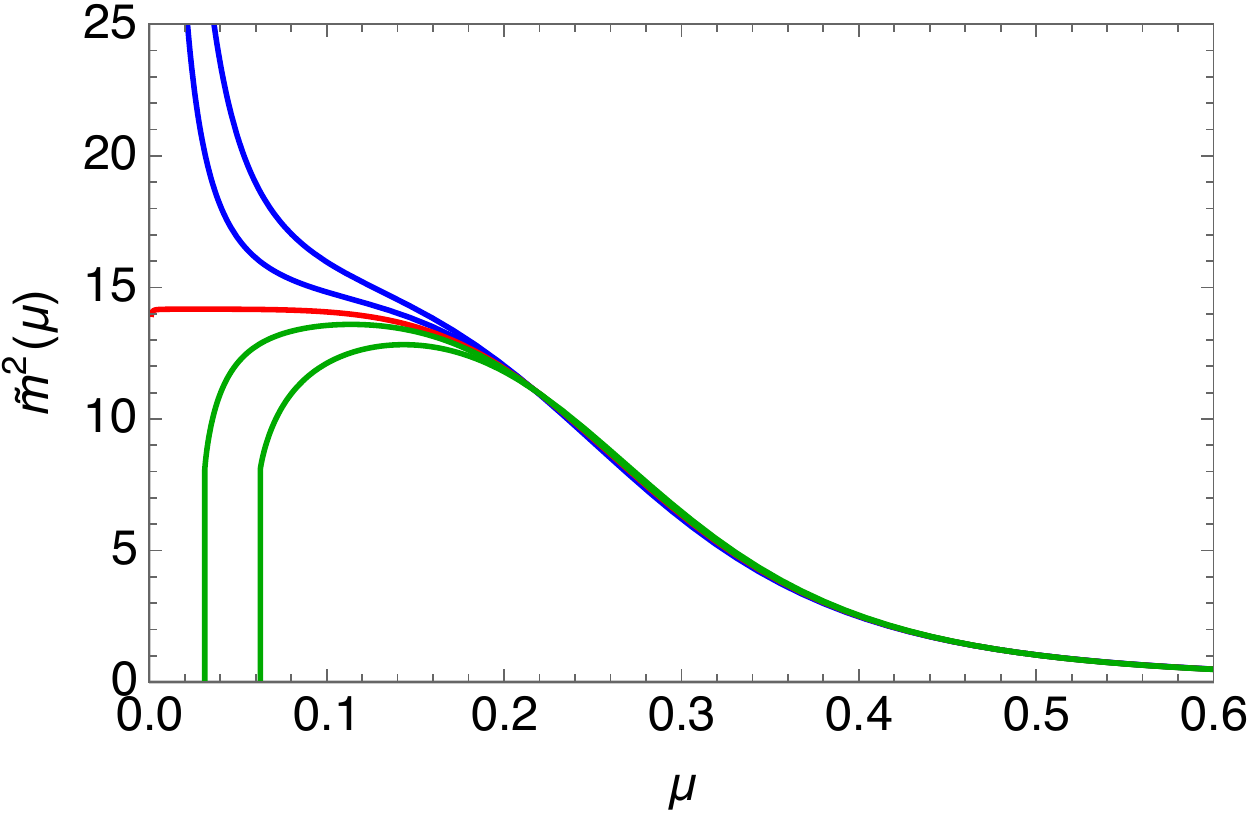}\\
  \includegraphics[width=.9\linewidth]{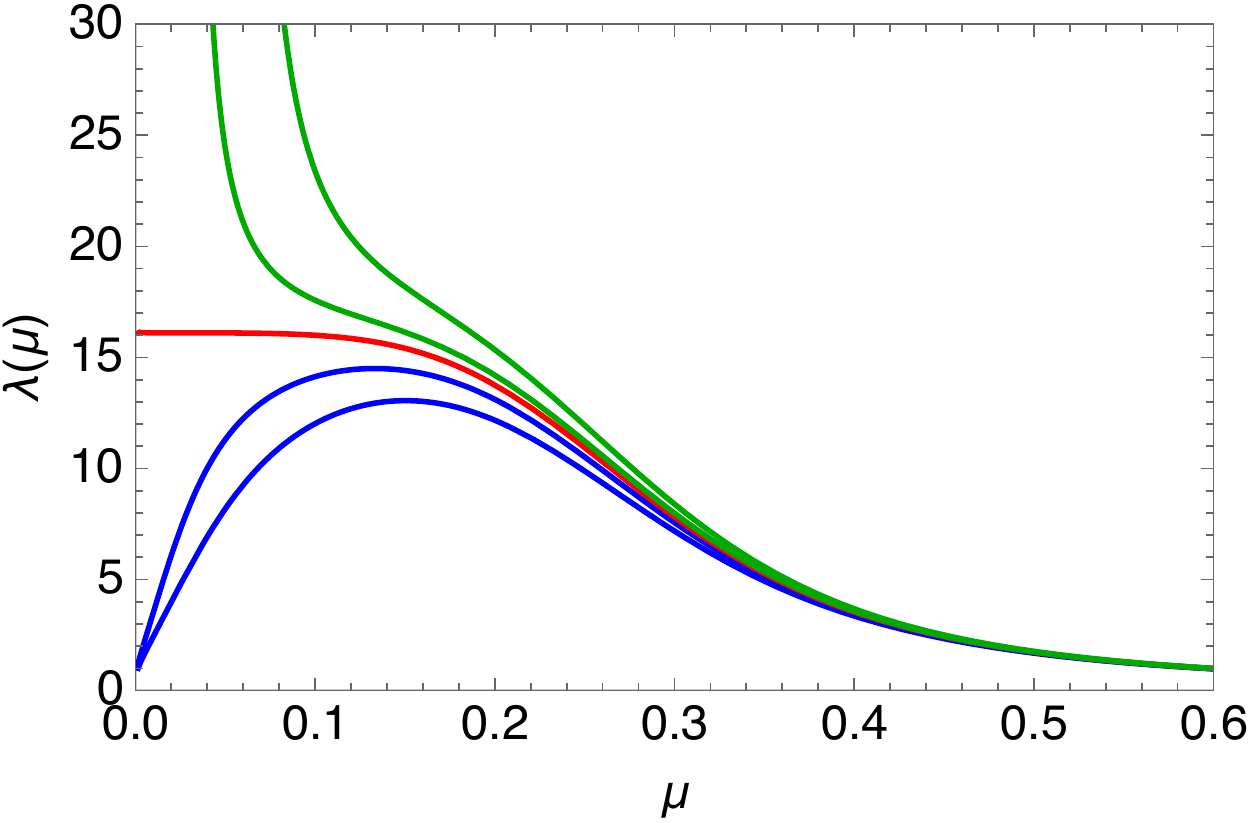}
  \caption{ The one-loop flow ($d=4$) of the reduced mass squared $\tilde m^2(\mu)$ (top) and of the coupling $\lambda(\mu)$ (bottom) for fixed $\lambda_0=3/\pi^2$ and various $m_0^2$ across the transition line (all in units of $\mu_0$). The red curve is the scaling solution (on the separatrix), corresponding to $m_0^2=m_{\rm scal}^2\approx0.27^2$, which runs into a fixed point. Green curves correspond to $m_0^2<m_{\rm scal}^2$ and have a Landau pole where the coupling diverges and the flow ends. Blue curves, with $m_0^2>m_{\rm scal}^2$, yield infrared-safe trajectories which correspond to decoupling solutions.}
  \label{fig_flow2}
\end{figure}

The UV behavior ($t\gg1$) is given by $\gamma_c\approx-3\lambda/2$ and $\gamma_A\approx-13\lambda/3$, from which we obtain the standard universal beta function of the coupling $\beta_\lambda\approx-22\lambda^2/3$ as well as $\beta_{m^2}/m^2\approx(35/44)\beta_\lambda/\lambda$. We thus recover the correct asymptotic behavior for the coupling $\lambda\sim3/[22\ln(\mu/\Lambda_{\rm L})]$, with $\Lambda_{\rm L}$ the scale of the perturbative Landau pole in the massless FP theory, and the mass $m^2\propto\lambda^{35/44}$ runs logarithmically. It follows that, in the UV, all trajectories in the plane $(\tilde m^2,\lambda)$ take the form 
\eq{\tilde m^2\propto \lambda^{35/44}e^{-3/(11\lambda)},}
and condense on the vertical axis, corresponding to the massless (FP) theory. There is an exponential focusing effect. 

The one-loop RG chart of the theory has been described in Ref.~\cite{Serreau:2012cg} and is shown in \Fig{fig_flow} in the plane $(\tilde m^2,\lambda)$. Focusing on regions which are connected to the Gaussian UV fixed point, there are two distinct phases on each side of a separatrix. In one phase, the flow towards the infrared ends at a Landau pole, where the coupling diverges at a nonzero $\mu$. The other phase contains infrared-safe trajectories, where the coupling and the mass stay finite all the way to the deep infrared and actually vanish logarithmically for $\mu\to0$; see below. Finally, the separatrix between these two phases relates the Gaussian UV fixed point to a nontrivial infrared fixed point. We show the RG running of both $\tilde m^2$ and $\lambda$ along the separatrix as well as typical flows on each side of it in \Fig{fig_flow2}. It is worth emphasizing that RG trajectories cannot cross the line $\tilde m^2=0$ because the beta function $\beta_{\tilde m^2}$ vanishes identically there. The same is true for the line $\lambda=0$. Moreover, the fact that $Z_A$, $Z_c$, and $g_B^2$ should all be positive or zero together with the first condition \eqn{eq:renpres2} imply that $\lambda_0\ge0$. Finally, the one-loop expressions of the RG functions only make sense for $\tilde m^2\ge0$. In the following, we thus restrict our attention to the quadrant $\tilde m^2\ge0$ and $\lambda\ge 0$.

\section{The infrared-safe phase}
\label{sec:IRphase}

We now discuss the form of the ghost and gluon propagators in the infrared-safe phase and on the separatrix and relate them to decoupling and scaling solutions. The ghost dressing function and the gluon propagator are completely determined by the values of the parameters $m_0^2$ and $\lambda_0$ at a given scale $\mu_0$. The expressions \eqn{eq:ghostprop} and \eqn{eq:glueprop} clearly show that different sets of parameters $(m_0^2,\lambda_0)$ belonging to a given RG trajectory yield the same functions $F(p)$ and $G(p)$ up to the overall normalizations $m^2_0/\lambda_0$ and $\lambda_0/m^4_0$, respectively. This merely states that the ghost and gluon propagators are RG invariant up to some normalization. 

Accordingly, in order to study the physically relevant parameter space of the theory, it is sufficient to choose units such that $\mu_0=1$ and consider one representative of each (relevant) RG trajectory (i.e., those connected to the Gaussian UV fixed point). In the following, we analyze the effect of varying the gluon mass parameter $m_0^2$ at a fixed value of the coupling $\lambda_0$. As we see from \Fig{fig_flow2}, at one-loop order, this does not intersect all possible (physically relevant) RG trajectories since the running coupling $\lambda(\mu)$ actually reaches a maximum value along infrared-safe trajectories.\footnote{To intersect all infrared-safe trajectories, one could, instead, fix the mass and vary the coupling. Note that, for fixed $\lambda_0$, there are two values of $\tilde m_0^2$ corresponding to a same trajectory. In the following, we consider intervals of $\tilde m_0^2$ so as to intersect each trajectory only once.} It is always possible to reach the region of parameter space corresponding to these trajectories by lowering the initial coupling, and we do not expect that the result presented below will be qualitatively affected. We thus choose a representative value for which the cases of practical interest (in particular, the parameter describing lattice results) are represented. In practice, we take $\lambda_0=3/\pi^2$, corresponding to $g_0=4$ in the SU($3$) theory in $d=4$.

\subsection{Decoupling solutions}

At one-loop order, the flow described by Eqs.~\eqn{m2IRscheme} and \eqn{eq:betatildem2}, with the gamma functions \eqn{eq:gammac} and \eqn{eq:gammaA}, has an attractive infrared fixed point located at $\lambda=1/\tilde m^2=0$. Generic infrared-safe trajectories are rapidly pushed towards $\tilde m^2\gg1$ and eventually flow to this weakly coupled fixed point. In this massive regime, we have $\gamma_A\approx \lambda/3$ and $\gamma_c\approx0$, from which it follows that 
\eq{\label{eq:behavior}
m^2(\mu)\propto\lambda(\mu)\sim 3/\ln(\bar\mu/\mu),
}
where $\bar\mu$ is an arbitrary scale. Using the expressions \eqn{eq:ghostprop} and \eqn{eq:glueprop} of the ghost dressing function and the gluon propagator, we conclude that, for generic initial conditions in the infrared-safe phase,  
\eq{\label{eq:decouplingtraj}
F(p\to0)\sim{\rm const}\quad{\rm and}\quad G(p\to0)\sim{\rm const}.
}
Here, the two constants depend on the initial condition $m^2_0$ and are related by the exact identity\footnote{This follows, with the present renormalization prescriptions, from a similar relation among bare quantities, which derives from the modified Slavnov-Taylor identities of the massive model and the assumption that the gluon two-point vertex function is  regular at $p=0$; see, e.g., \cite{Tissier:2011ey}. The latter is satisfied for the infrared-safe trajectories described here.} 
\eq{\label{eq:relationmass}
G(0)F(0)=1/m_0^2.
} 
 Equation \eqn{eq:decouplingtraj} describes a decoupling solution. Notice that the RG flow drives the system towards a weak coupling regime, which justifies {\it a posteriori} the use of the one-loop approximation.

Furthermore, the behavior \eqn{eq:behavior} allows us to describe the shape of the infrared-safe trajectories in the plane $(\tilde m^2,\lambda)$ in the deep infrared. We have 
\eq{\label{eq:phaseboundary}
\tilde m^2\propto\lambda/\mu^2\propto\lambda\,e^{6/\lambda}.
}
In particular, there exists a value of the proportionality coefficient corresponding to the curve limiting the infrared-safe phase in the region of large $\tilde m^2$. This is another separatrix of the flow, which is visible in dashed line on \Fig{fig_flow}. Trajectories beyond that line are not connected to the Gaussian UV fixed point. 

Finally, it is interesting to compute the following running coupling:
\eq{\label{eq:conventionalTaylor}
\lambda_T(p)=\lambda_0p^2G(p)F^2(p)=\frac{\lambda(p)}{1+\tilde m^2(p)},
}
which is used in both lattice \cite{Cucchieri:2004sq,Ilgenfritz:2006gp,Bogolubsky09} and continuum \cite{Lerche:2002ep,Schleifenbaum:2006bq,Fischer08} studies. It corresponds to the standard Taylor coupling in the massless FP theory, in which case, it is simply identical to the coupling $\lambda$. Also, as argued in Ref.~\cite{Tissier:2011ey}, $\lambda_T(p)$ is the relevant loop-expansion parameter, both in the massless ($\tilde m^2\ll1$) and in the massive ($\tilde m^2\gg1$) regimes. 
For the infrared-safe trajectories \eqn{eq:behavior}, we have the power law behavior 
\eq{\label{eq:conventionalTaylordec}
\lambda_T(p\to0)\sim\frac{\lambda_0F(0)}{m_0^2} p^2,
}
in agreement with lattice results.

\subsection{The scaling solution}

The flow may also have a fixed point at nonzero, finite values $\tilde m_*^2$ and $\tilde\lambda_*$ of the parameters. Demanding $\beta_{\tilde m^2}/\tilde m^2=\beta_\lambda/\lambda=0$ in Eqs.~\eqn{m2IRscheme} and \eqn{eq:betatildem2} yields 
\eq{\label{eq:anomalousdim}\gamma_c^*=-2\,,\quad\gamma_A^*=4.}
At one-loop order Eqs.~\eqn{eq:gammac} and \eqn{eq:gammaA} give a nontrivial solution with\footnote{Approximate values can be obtained from the approximate gamma functions in the massive ($t\ll1$) regime. This is because $t_*=1/\tilde m_*^2\approx0.07$.  Using the low-$t$ expansions
\eq{
\gamma_A&=\frac{\lambda}{3}-\frac{217\lambda t}{180}+{\cal O}(t^2)\quad{\rm and}\quad\gamma_c=\frac{\lambda t}{2}\left(\ln t-\frac{5}{6}\right)+{\cal O}(t^2),\nonumber
}
we find
$$\ln \tilde m_*^2\approx\frac{\tilde m_*^2}{3}-\frac{367}{180}\quad{\rm and}\quad
\tilde\lambda_*\approx\frac{720\tilde m_*^2}{60\tilde m_*^2-217}.
$$
This gives the approximate values $\lambda_*\approx16.16$ and $\tilde m_*^2\approx14.04$.
}
\eq{
\label{eq:scalFPd4}
\tilde\lambda_*\approx 16.11\,,\quad \tilde m_*^2\approx 14.18,
}
as observed in \Fig{fig_flow}.
This corresponds to $g_*=35.66$ for $N=2$ and to $g_*\approx29.12$ for $N=3$. Clearly, this fixed point solution corresponds to a strong coupling regime for which the one-loop analysis is questionable. We note, however, that the situation is less dramatic than the large value of $\tilde\lambda_*$ indicates when measured in terms of the relevant expansion parameter in the infrared $\lambda_T\to\tilde\lambda_{T}^*$, that is
\begin{figure}[t!]
  \centering
  \includegraphics[width=.9\linewidth]{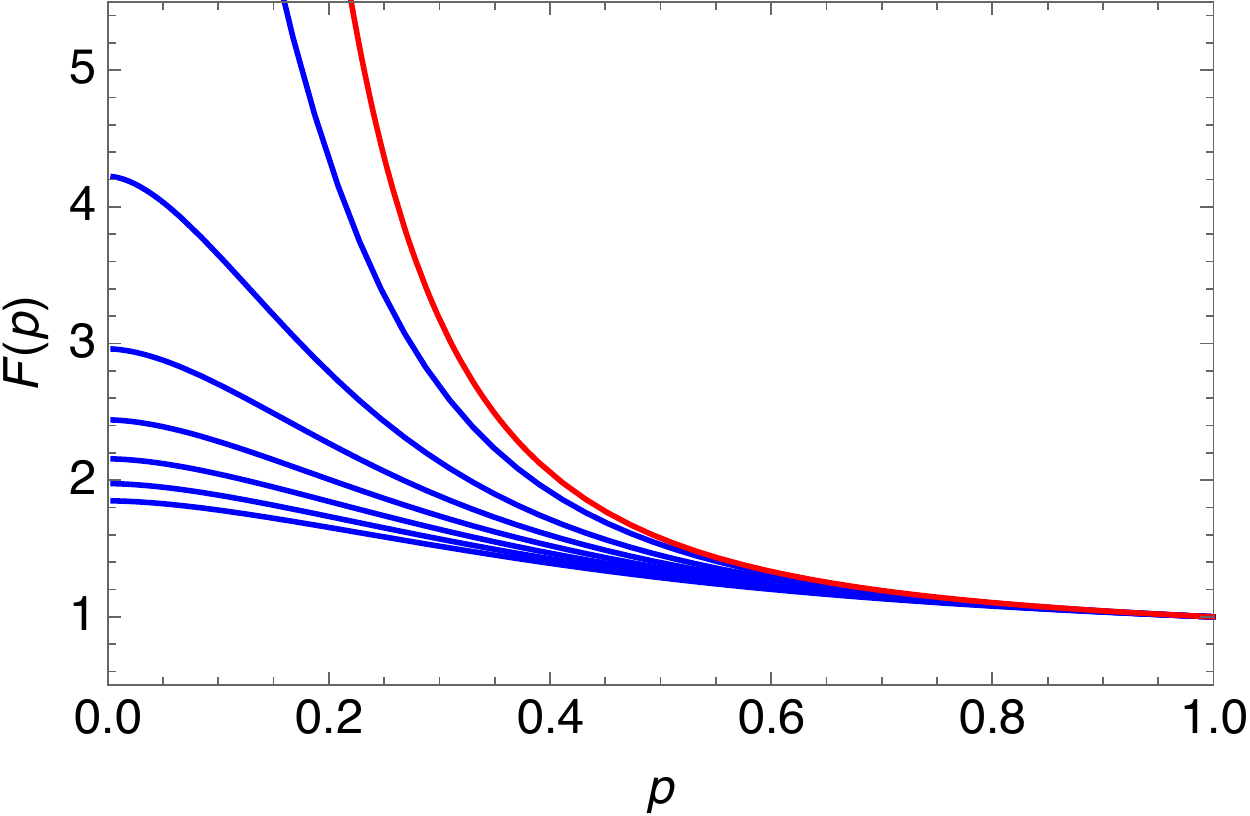}
  \includegraphics[width=.9\linewidth]{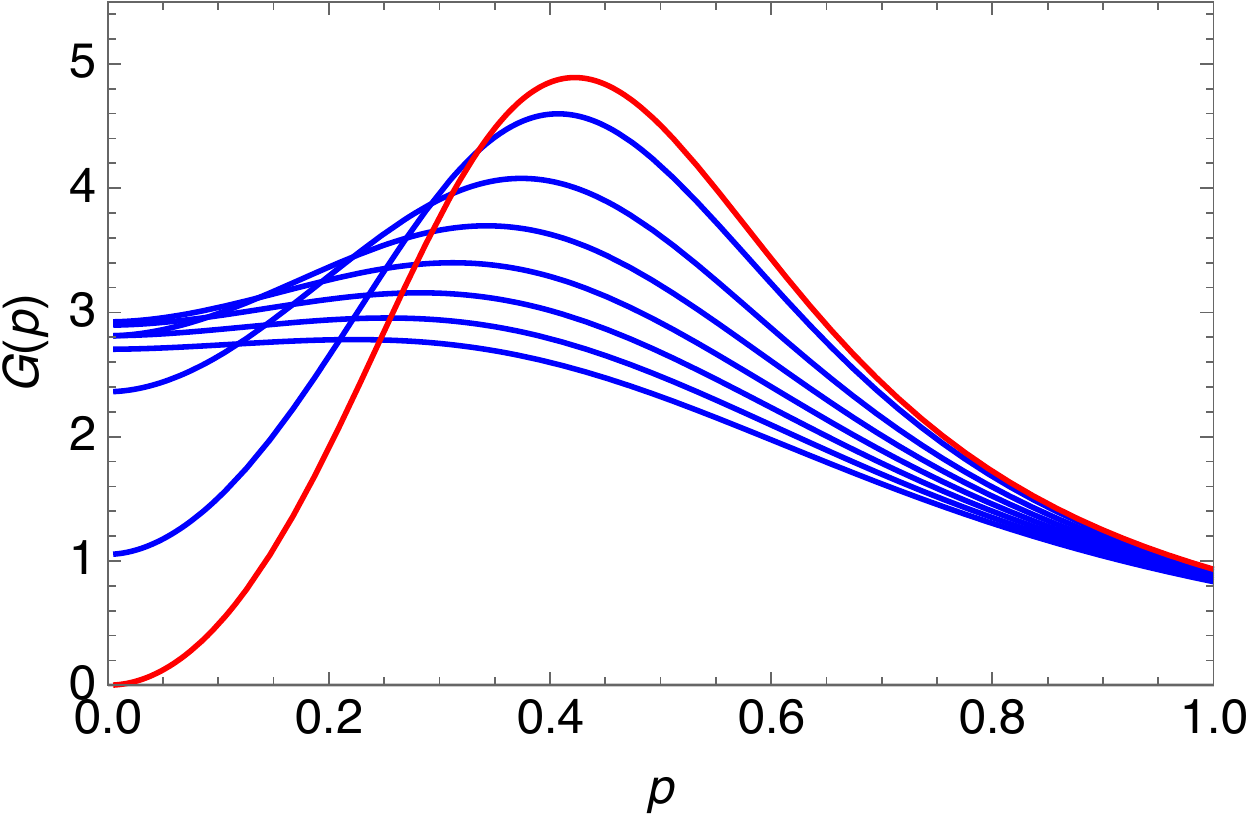}\\
  \caption{The ghost dressing function (top) and gluon propagator (bottom) at one-loop order as functions of momentum (units of $\mu_0$) for $\lambda_0=3/\pi^2$ and various $m_0^2$ from the critical value $m_0^2=m_{\rm scal}^2\approx0.27^2$, corresponding to the scaling solution (red), to deeper in the infrared-safe phase, up to $m_0^2=0.45^2$, describing decoupling solutions (blue curves).}
  \label{fig_propag}
\end{figure}
\eq{\label{eq:conventionalTaylorvalue}
\tilde\lambda_{T}^*=\frac{\tilde\lambda_*}{1+\tilde m_*^2}\approx 1.06.
}

Furthermore, we stress that one can infer nontrivial information on the solutions reachable by our perturbative approach solely based on the assumption of the existence of an infrared fixed point at finite, nonzero $\tilde m^2_*$ and $\lambda_*$. For instance, the values \eqn{eq:anomalousdim} of the anomalous dimensions only assume that such a fixed point exists and do not rely on the one-loop approximation. Moreover, the infrared behaviors of the ghost and gluon propagators on the critical trajectory ending at the fixed point is of the scaling type, with exponents entirely determined by the anomalous dimensions \eqn{eq:anomalousdim}. Adjusting $\tilde m_0^2=\tilde m^2_{\rm sep}(\lambda_0)$, with $\tilde m^2_{\rm sep}(\lambda)$ the equation of the separatrix, we get, from Eqs.~\eqn{eq:ghostprop} and \eqn{eq:glueprop},
\eq{\label{eq:scalingghost}
F(p\to0)&\sim\frac{m^2_0}{\lambda_0}\frac{\tilde\lambda_*}{\tilde m^2_*} p^{-2},\\
\label{eq:scalinggluon}
G(p\to0)&\sim\frac{\lambda_0}{m^4_0}\frac{\tilde m^4_*}{\tilde\lambda_*}\frac{p^2}{1+\tilde m^2_*} .
}
The scaling exponents, defined as 
\eq{\label{eq:scalingexpdef}
p^2G(p)\sim p^{2\alpha_G}\quad {\rm and}\quad F(p)\sim p^{2\alpha_F},
}
are related to the anomalous dimensions as $\alpha_G=\gamma_A^*/2$ and $\alpha_F=\gamma_c^*/2$. Our scaling solution is of the Gribov type \cite{Gribov77}, with
\eq{
\alpha_G=-2\alpha_F=2.
}

Finally, we also note that, because the inverse gluon propagator is nonanalytic at $p\to0$, the relation \eqn{eq:relationmass} does not hold. Instead, we have 
\eq{
m^2_0G(p)F(p)|_{p\to0}=\frac{\tilde m_*^2}{1+\tilde m_*^2}\approx1,
} 
where we have used the one-loop result only in the final estimation. We plot the ghost dressing function and the gluon propagator for $d=4$ in \Fig{fig_propag}. The different curves are obtained by integrating the one-loop RG flow for various initial parameters, from the separatrix (scaling solution) to deeper in the infrared-safe phase (decoupling solutions).  The structure of the space of solutions, with continuous families of singular versus decoupling solutions on each side of a scaling solution is reminiscent of what is observed in studies based on nonperturbative continuum approaches; see, e.g., Ref.~\cite{Cyrol:2016tym}.

\section{General dimension $2\le d\le4$}

It is interesting to generalize the previous discussions to arbitrary dimensions. We introduce the dimensionless coupling 
\eq{
\tilde\lambda=\mu^{d-4}\frac{g^2N}{(4\pi)^{d/2}\Gamma(d/2)}
} and we work with the variable $t=1/\tilde m^2$, which proves more convenient to describe the infrared massive regime $t\ll1$. The corresponding beta functions read
\eq{
\label{eq:betam2d}\beta_t=t(2-\gamma_A-\gamma_c)
}
and 
\eq{
\label{eq:betald}\beta_{\tilde\lambda}=\tilde\lambda(d-4+\gamma_A+2\gamma_c),
}
where the functions $\gamma_{A,c}\equiv\gamma_{A,c}(t,\tilde\lambda)$ depend on the dimension. Their one-loop expressions have been derived in the integer dimensions $d=2,3,4$ in Ref.~\cite{Tissier:2011ey}, and their generalizations to arbitrary $d$ can be deduced from the material presented in that reference; see also Ref.~\cite{Reinosa:2013twa}. They involve combinations of hypergeometric functions which we shall exploit numerically below but which are not particularly enlightening. However, they assume simpler forms in the massive regime $t\ll1$, which will be of interest for our purposes in this section. Generalizing the analysis of \cite{Tissier:2011ey}, we get
\eq{
\label{eq:gamAd}\gamma_A&= X(d)\tilde\lambda+{\cal O}(\tilde\lambda t^{6-d\over2})
}
and
\eq{
\label{eq:gamcd}\gamma_c=-\frac{d-1}{4-d}X(d)\tilde\lambda t+{\cal O}(\tilde\lambda t^{6-d\over2}),
}
with 
\eq{
\label{eq:moiaussi}
X(d)=\frac{2\Gamma^3(d/2)\Gamma(3-d/2)}{\Gamma(d)}.
}
The neglected terms in the expansion \eqn{eq:gamAd} and \eqn{eq:gamcd} are subleading for $d<4$ and the lower the dimension, the better the approximation. We also introduce the dimensionless rescaled coupling 
\eq{\label{eq:rescaledtilde}
\tilde\lambda_T=\frac{\tilde\lambda}{1+\tilde m^2}=\frac{\tilde\lambda \,t}{1+t},
}
which, again, is the relevant expansion parameter over the whole momentum range \cite{Tissier:2011ey}.
Finally, we note that the exponents defined in \Eqn{eq:scalingexpdef} can be obtained as
\eq{
\label{eq:alphaFd}2\alpha_F&=\left.\frac{d\ln F(p)}{d\ln p}\right|_{p\to0}=\gamma_c|_{p\to0}\\
\label{eq:alphaGd}2\alpha_G&=\left.\frac{d\ln p^2G(p)}{d\ln p}\right|_{p\to0}=\left.\frac{2-\gamma_c+\gamma_At}{1+t}\right|_{p\to0}.
}

We shall first discuss the two infrared fixed points corresponding to the decoupling and scaling solutions. We then analyze in detail the case $d\to2$, where both fixed points are at weak coupling and which can, thus, be described perturbatively.

\subsection{The decoupling fixed point}

The flow \eqn{eq:betam2d}--\eqn{eq:moiaussi} has a fixed point at 
\eq{
\label{eq:decFPd}
\tilde\lambda_*^{\rm dec}=\frac{4-d}{X(d)}\quad{\rm and}\quad t_*^{\rm dec}=0.
}
Note that $\tilde\lambda_*^{\rm dec}>0$ for $d<4$. A simple stability analysis shows that the eigenvalues of the linearized flow are $4-d$ and $d-2$, so that the decoupling fixed point \eqn{eq:decFPd} is infrared stable\footnote{One of the eigendirections is the inverse square mass $t\sim \mu^{d-2}$. We thus verify that the infrared flow is, indeed, driven towards the massive regime $t\ll1$ for $d>2$. The case $d=2$ is discussed below. [Note that, in the present scheme, the dimensionful mass squared, which vanishes as  $m^2\propto \mu^{4-d}$ in the infrared, differs from the screening mass squared $G^{-1}(p=0)$, which tends to a finite, nonzero value.]} for $2<d<4$. It corresponds to $\gamma_c^*=0$ and $\gamma_A^*=4-d$ and, in turn, from Eqs.~\eqn{eq:alphaFd} and \eqn{eq:alphaGd}, to a decoupling solution
\eq{
\label{eq:decouplingexpd}
\alpha_F=0\,,\quad\alpha_G=1.
}
Finally, we note that, in terms of the rescaled coupling \eqn{eq:rescaledtilde}, the decoupling fixed point is at $\tilde\lambda_{T*}^{\rm dec}=0$. In the infrared regime, we have 
\eq{\label{eq:FPlT}
\tilde \lambda_T(p)\propto p^{d-2}
} 
and the corresponding dimensionful coupling $\lambda_T=p^{4-d}\tilde\lambda_T\propto p^2$ in all dimensions. In particular, this justifies the present one-loop analysis for $d>2$.

\subsection{The scaling fixed point}

\begin{figure}[t!]
  \centering
  \includegraphics[width=.9\linewidth]{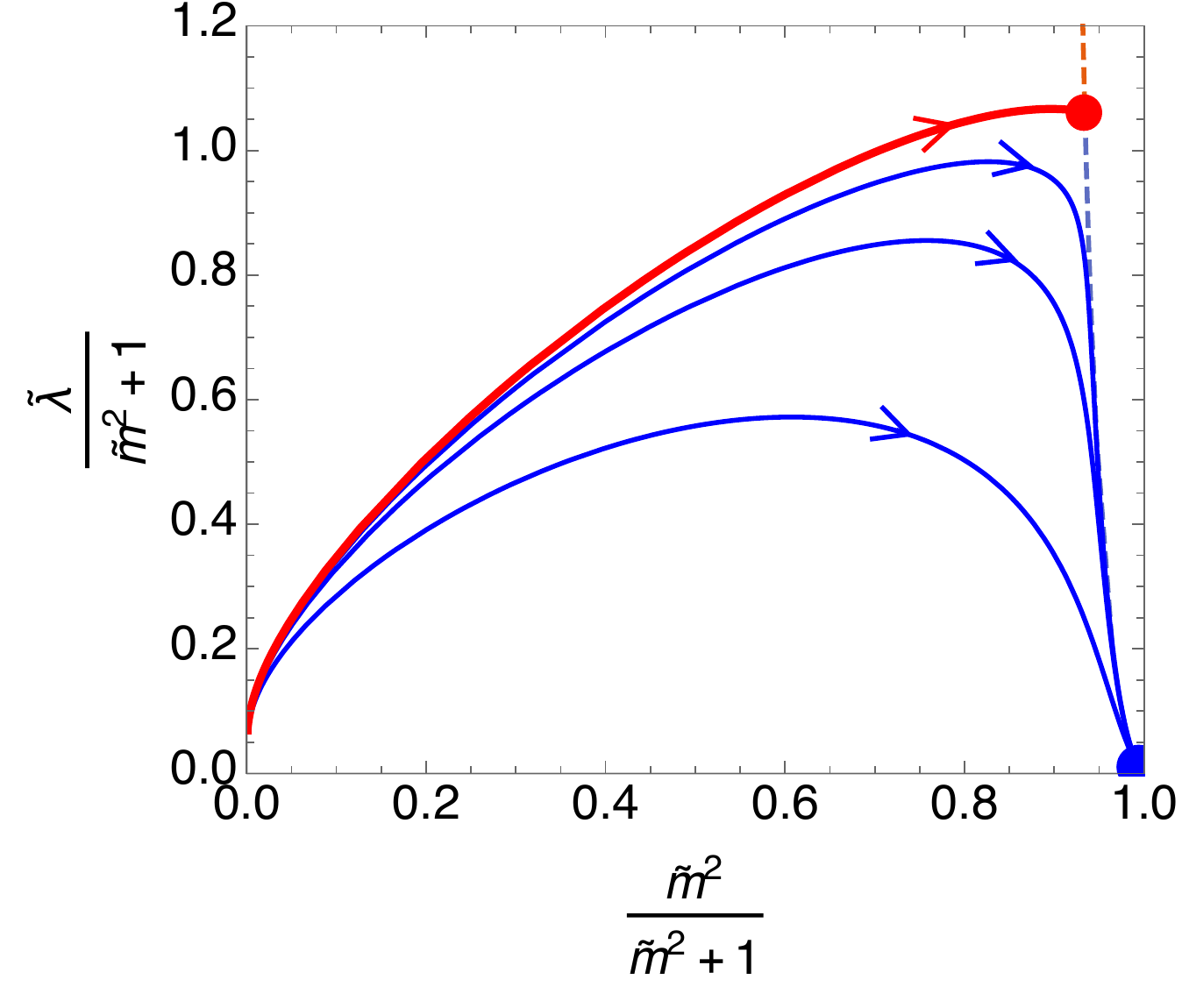}
  \caption{ The RG flow of \Fig{fig_flow} ($d=4$) in the rescaled variables $\tilde m^2/(1+\tilde m^2)$ and $\tilde\lambda_T=\tilde\lambda/(1+\tilde m^2)$. We only show the separatrix and some infrared-safe trajectories. The scaling and decoupling fixed point are represented by the red and blue dots, respectively.}
  \label{fig_flow-rescaled}
\end{figure}

As in the $d=4$ case, other possible fixed points may exist at finite nonzero $\tilde m^2$ (or $t$) and $\tilde\lambda$. Imposing $\beta_t/t=\beta_{\tilde\lambda}/\tilde\lambda=0$ in Eqs.~\eqn{eq:betam2d} and \eqn{eq:betald} implies the anomalous dimensions $\gamma_c^*=2-d$ and $\gamma_A^*=d$.  In turn, Eqs.~\eqn{eq:alphaFd} and \eqn{eq:alphaGd} yield the scaling exponents
\eq{\label{eq:scalingexpd}
\alpha_F=\frac{2-d}{2}\quad {\rm and}\quad\alpha_G=\frac{d}{2}.
}
These obviously satisfy the scaling relation $\alpha_G+2\alpha_F=(d-4)/2$, which follows from the fixed point equation $\beta_{\tilde\lambda}=0$. It is important to remark that the requirement of a nontrivial solution to $\beta_t=0$ implies the further constraint $\alpha_G+\alpha_F=1$ in the present renormalization scheme. As before, we emphasize that the scaling behavior \eqn{eq:scalingexpd} is an all-order statement which only relies on the assumption of a nontrivial infrared fixed point at $0<m^2_*,\tilde\lambda_*<\infty$. Using the one-loop expressions of the functions $\gamma_A$ and $\gamma_c$ in $d=3$ derived in Ref.~\cite{Tissier:2011ey}, one can check explicitly that the structure of the RG flow in $d=3$ is similar to the one described above in $d=4$, with distinct Landau pole and infrared-safe phases and a separatrix joining the ultraviolet Gaussian fixed point to the scaling fixed point. The latter is located at [for $d=3$, we have $\lambda=g^2N/(4\pi^2)$] 
\eq{
\label{eq:scalFPd3}
\lambda_*^{d=3}\approx5.48\quad {\rm and }\quad \left(\tilde m_*^2\right)^{d=3}\approx 4.78,
}
for which $\tilde\lambda_{T*}^{d=3}\approx0.95$. As for $d=4$, the scaling fixed point is at relatively strong coupling and the one-loop approximation is questionable at the quantitative level.

\begin{figure}[t!]
  \centering
  \vspace{.15cm}
  \includegraphics[width=.9\linewidth]{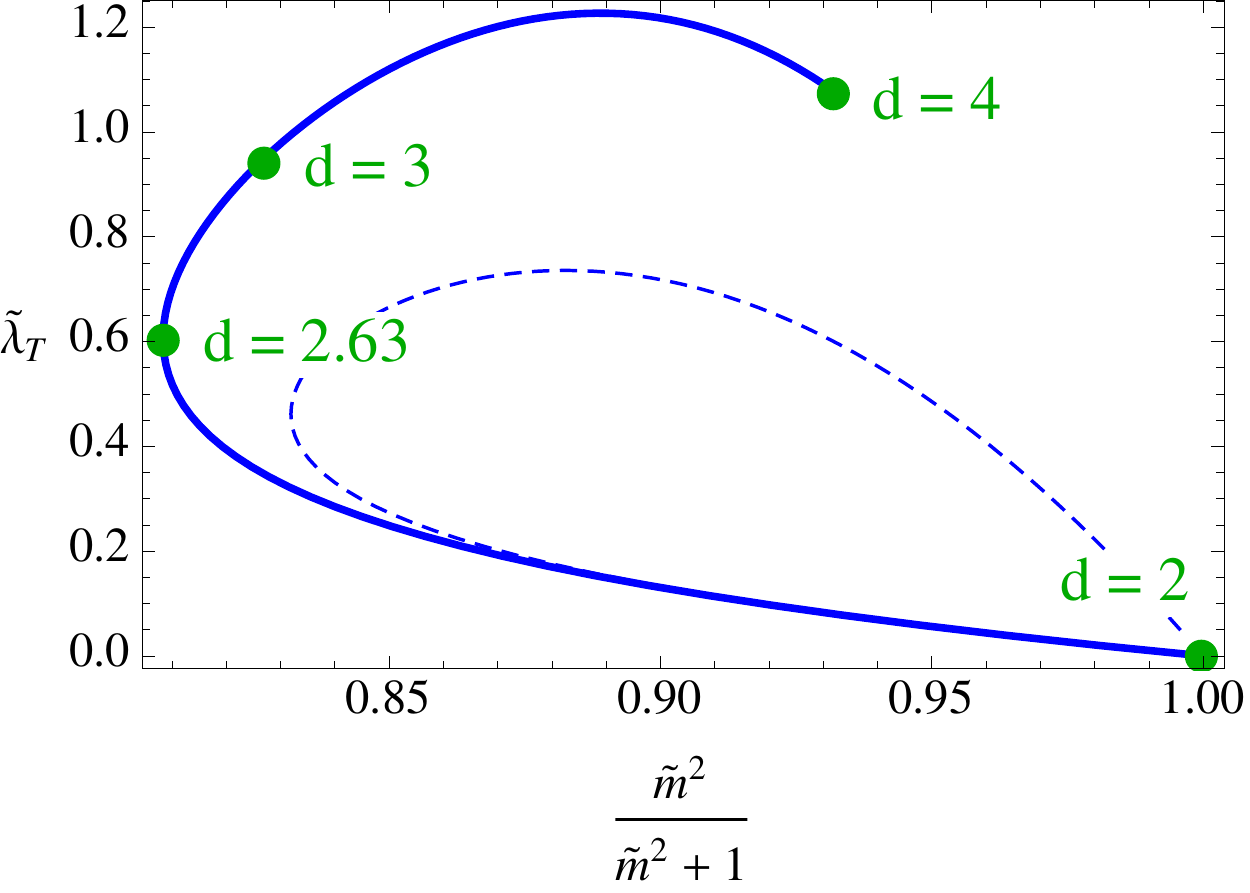}
  \caption{ Evolution of the scaling fixed point in the plane $(\tilde m^2/(1+\tilde m^2),\tilde\lambda_T=\tilde\lambda/(1+\tilde m^2))$ as the dimension $d$ is varied from $4$ to $2$. The dashed curve corresponds to the approximated formula \eqn{eq:scalFPdd}. For $d\to2$, the scaling fixed point merges with the decoupling one at $\tilde\lambda_T=1/\tilde m^2=0$.}
  \label{d_fp}
\end{figure}

We can study the general case $2\le d\le4$ by evaluating the appropriate hypergeometric functions mentioned earlier. For $d\neq2$, we find the same structure as before, with a scaling fixed point at $0<\tilde m^2_*,\tilde\lambda_*<\infty$. For $d=2$, the latter merges with the decoupling fixed point at $\tilde m^2_*=\infty$, i.e., at $\tilde\lambda_{T*}=0$ (see also the discussion in the next subsection). This is represented on Figs. \ref{fig_flow-rescaled} and \ref{d_fp}. As we decrease $d$, the value of $\tilde m_*^2$ first decreases and then increases again towards arbitrarily large values, with a turning point at a dimension $d\approx2.63$. In this regime, the expressions \eqn{eq:gamAd} and \eqn{eq:gamcd} provide good approximations from which we can get a simple analytic control. These give
\eq{
\label{eq:scalFPdd}
\tilde\lambda_*^{\rm scal}\approx\frac{d}{X(d)}\quad{\rm and}\quad t_*^{\rm scal}\approx\frac{(4-d)(d-2)}{d(d-1)}.
}
As expected, this is not a good description of the $d=4$ fixed point \eqn{eq:scalFPd4} despite the fact that the latter sits at a relatively large $\tilde m_*^2$. This is because the expansion \eqn{eq:gamAd}--\eqn{eq:gamcd} is not valid in that case due to large logarithmic corrections. For $d=3$, although the square mass at the fixed point \eqn{eq:scalFPd3} is not very large, we get the qualitatively good estimate $\tilde\lambda_*^{\rm scal}\approx3/X(3)=48/\pi^2\approx4.86$ and $ t_*^{\rm scal}\approx1/6$. Obviously, this gets better as one decreases the dimension, as demonstrated in \Fig{d_fp}. Finally, the massive approximation allows us to analyze the stability of the fixed point. A simple calculation shows that the eigenvalues of the linearized flow around the fixed point \eqn{eq:scalFPdd} are given by the anomalous dimensions $\gamma_A^*=d$ and $\gamma_c^*=2-d$, which have opposite signs for $d>2$: The scaling fixed point has one unstable direction.

The present analysis shows that, in terms of the coupling $\tilde\lambda_T$, the scaling fixed point becomes perturbative for $d\to2$, with $\tilde\lambda_T^*\approx\tilde\lambda_* t_*\to d-2$. In the next subsection, we discuss this limit in detail and we make a link with the analysis of Ref.~\cite{Weber:2011nw}.

\subsection{The case $d=2+\epsilon$}
\label{appsec:2d}

In that case, both the decoupling and the scaling fixed points are well described by the massive regime of the RG flow. From the previous discussions, we get, up to ${\cal O}(\epsilon^2)$ corrections, 
\eq{
\tilde\lambda_*^{\rm dec}=1+\epsilon\,,\quad  t_*^{\rm dec}=0
}
and 
\eq{
\tilde\lambda_*^{\rm scal}=1+2\epsilon\,,\quad t_*^{\rm scal}=\epsilon.
}
The eigenvalues of the linearized flow are $2-\epsilon$ and $\epsilon$ for the decoupling fixed point and $2+\epsilon$ and $-\epsilon$ for the scaling fixed point.

An equivalent, more appropriate description of the flow near these fixed points can be made in terms of the variable $\tilde\lambda_T$. Note that, in the massive regime, the latter reduces to $\tilde\lambda_T\approx\tilde\lambda/\tilde m^2$, which has also been introduced in \cite{Tissier:2011ey,Weber:2011nw}. We have, up to ${\cal O}(\epsilon)$ relative corrections, 
\eq{
\beta_{\tilde\lambda_T}&=\tilde\lambda_T(\epsilon-\tilde\lambda_T)\\
\beta_t&=2(t-\tilde\lambda_T).
}
The first equation coincides with the one derived by Weber in Ref.~\cite{Weber:2011nw}, though in a different scheme, where, in particular, the mass $m^2$ does not run in the infrared. But, as we see here, the flow of $\tilde\lambda_T$ is independent of the mass in the limit $\epsilon \to 0$. The two fixed points are 
\eq{
\tilde\lambda_{T*}^{\rm dec}&=t_*^{\rm dec}=0\quad{\rm (decoupling, IR\, stable)}\\
\label{eq:lambdatildestar2d}\tilde\lambda_{T*}^{\rm scal}&=t_*^{\rm scal}=\epsilon\quad{\rm (scaling, IR\, unstable)}
}
They are related by a trajectory $\tilde\lambda_T=t$ and the infrared-safe trajectories are such that $\tilde\lambda_T\sim \mu^\epsilon$ and $\tilde\lambda_T-t\sim\mu^2$.

For $d=2$, the two fixed point merge at $\tilde\lambda_T=0$ and the decoupling and scaling exponents \eqn{eq:decouplingexpd} and \eqn{eq:scalingexpd} become identical. The resulting fixed point is unstable in the direction $\tilde\lambda_T$ and there is no infrared-safe phase left: All RG trajectories have a Landau pole, invalidating the perturbative analysis. 

To summarize, the present RG analysis at one-loop order gives a picture in qualitative agreement with the lattice results in $d=4$ and $d=3$, where the infrared stable fixed point corresponds to a decoupling solution. In $d=2$, there is no stable fixed point at one-loop order, which is also in line with the fact that lattice simulations do not find a decoupling or a Gribov-like scaling solution, but yield a different scaling behavior with non-Gribov exponents, as discussed in \Sec{sec:comparison} below. This suggests that the corresponding fixed point may either require another renormalization scheme or more general deformations of the FP Lagrangian, or it may be truly nonperturbative.

\section{Discussion}
\label{sec:discussion}

With the present perturbative approach, we are in a position to discuss in a simple way the properties of the decoupling and scaling solutions in relation with various issues raised in the literature. In particular, we consider here the relation between the ghost dressing function at zero momentum, the gluon screening mass squared, and the control mass parameter $m_0^2$. Another quantity of interest concerns the scale of spectral positivity violation and its dependence on $m_0^2$.  Finally, we analyze the longitudinal component of the gluon two-point vertex function and we discuss the scaling solution in relation with the issue of BRST symmetry restoration mentioned in the Introduction.

\subsection{Ghost dressing and gluon screening mass}

\begin{figure}[t!]
  \centering
  \includegraphics[width=.9\linewidth]{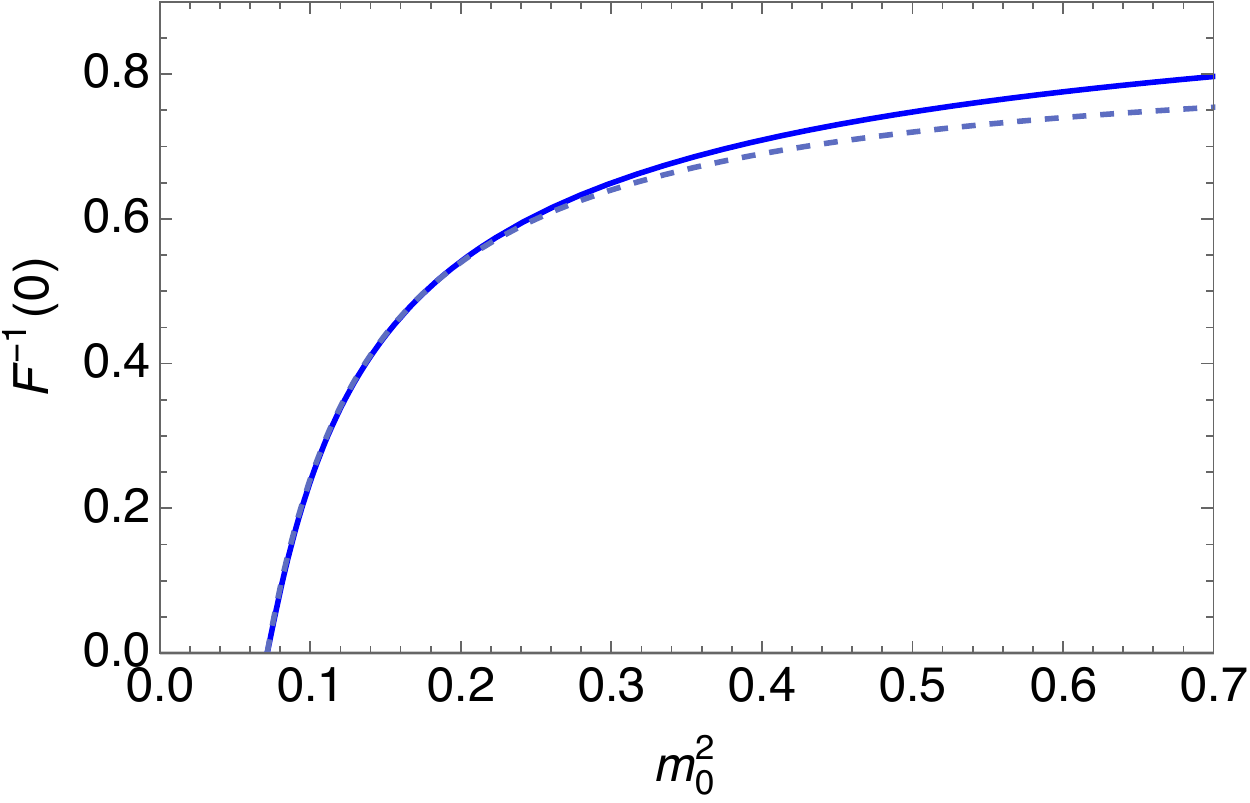}\\
  \includegraphics[width=.9\linewidth]{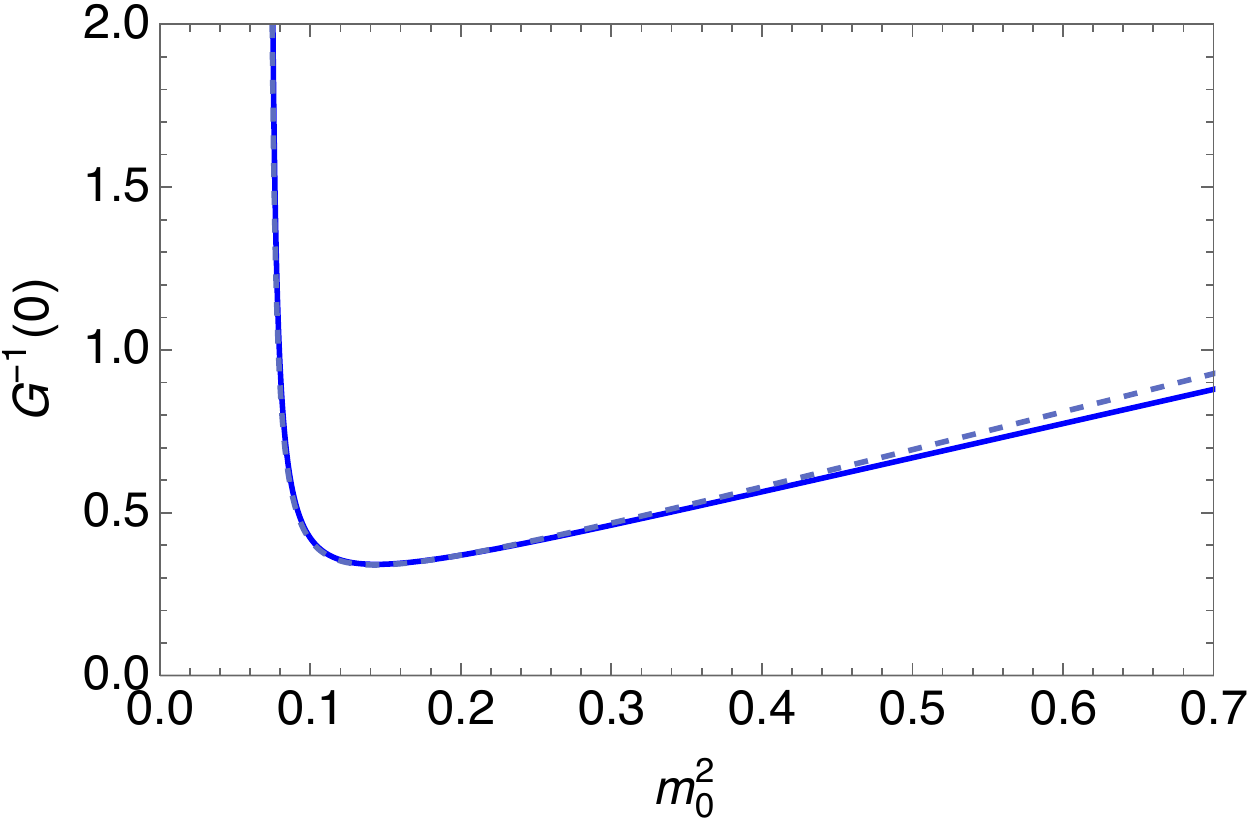}
  \caption{Top: the ghost wave function renormalization $F^{-1}(0)$ as a function of the mass parameter $m^2_0$ and fixed $\lambda_0=3/\pi^2$ (units of $\mu_0$ and $d=4$). Bottom: the gluon screening mass squared $G^{-1}(0)$ as a function of $m_0^2$ for the same parameters. The value $m_0^2=m_{\rm scal}^2\approx 0.27^2\mu_0^2=0.073\mu_0^2$ corresponds to the scaling solution. The dashed lines show the approximate behaviors \eqn{eq:approxghostwf} and \eqn{eq:screeningapprox}.}
  \label{fig_dressing}
\end{figure}

A first question of interest is the relation between the ghost wave function renormalization, given by the inverse dressing function at vanishing momentum $F^{-1}(0)$, and the gluon mass parameter $m_0^2$. The former appears as a boundary condition in DSE calculations. It plays the role of the control parameter for the family of decoupling solutions and takes the particular value $F^{-1}(0)=0$ for the scaling solution. Instead, $m_0^2$ is related, although in a nontrivial way\footnote{ The present gluon mass parameter $m_0^2$ and the one employed in FRG studies (see, e.g., Ref.~\cite{Cyrol:2016tym}) are defined in very different setups and renormalization schemes. Note, for instance, that the running square mass parameter employed in that reference receives quadratic contributions due to the explicitly BRST breaking regulator. As mentioned previously, in the present scheme, the running of the mass parameter is only logarithmic (in $d=4$), being protected by the BRST symmetry of the massless limit.}, to the control parameter of FRG studies. In the present scheme, we can easily compute $F^{-1}(0)$ as a function of $m_0^2$ for a given $\lambda_0$. This is shown in the upper panel of  \Fig{fig_dressing}. The curve starts at the scaling value $m_{\rm scal}^2=\mu_0^2\,\tilde m_{\rm sep}^2(\lambda_0)$, for which $F^{-1}(0)=0$, and rises monotonously with increasing values of $m_0^2$, thereby describing the whole family of decoupling solution. Using the infrared solution \eqn{eq:behavior} (or \Eqn{eq:FPlT} for arbitrary $d$), one easily concludes that $F^{-1}(0)$ approaches $1$ for large mass $m_0^2/\mu_0^2\gg1$.

Next, we consider the evolution of the gluon screening mass squared $G^{-1}(0)$ as a function of $m_0^2$. As already emphasized, in the present renormalization scheme, the identity \eqn{eq:relationmass} holds whenever the inverse gluon propagator is analytic at $p\to0$, which is the case for decoupling solutions. The screening mass squared is thus given by $G^{-1}(0)=m^2_0F(0)$ and its dependence on $m_0^2$, shown in the lower panel of \Fig{fig_dressing} is completely governed by that of the ghost wave function renormalization $F^{-1}(0)$ discussed above. First, we recover the fact that in the limit $m_0^2\to m_{\rm scal}^2$, where $F^{-1}(0)\to0$, the gluon screening mass diverges, corresponding to $G(0)\to0$ as discussed previously. Next, we see that $G^{-1}(0)$ presents a counterintuitive nonmonotonous behavior, first pointed out in Ref.~\cite{Cyrol:2016tym}, where it decreases for $m_0^2$ close to the scaling solution and then increases for larger $m_0^2$. 

The decrease at low $m_0^2-m_{\rm scal}^2$ is a direct consequence of the behavior of $F^{-1}(0)$ in this region. What happens in this regime can be understood as follows.
In the infrared limit, we have, from \Eqn{eq:behavior} for $d=4$ and from \Eqn{eq:FPlT} in general dimension, that  $m^2(p)/\lambda(p)\approx p^{d-2}/\tilde\lambda_T(p)\sim {\rm const}$. This constant being proportional to $F^{-1}(0)$, it must vanish on the separatrix. We thus expect $m^2(p)/\lambda(p)|_{p\to0}=c\left(m_0^2-m_{\rm scal}^2\right)$, where both $c$ and $m_{\rm scal}^2$ depend on $\lambda_0$. We have checked that this is indeed the case, and we get $c\approx 2.766$ for $\lambda_0=3/\pi^2$ and $d=4$. It follows that
\eq{\label{eq:approxghostwf}
F^{-1}(0)\sim\frac{c\lambda_0}{m_0^2}\left(m_0^2-m_{\rm scal}^2\right)
}
and, thus,
\eq{\label{eq:screeningapprox}
G^{-1}(0)\sim\frac{m_0^4}{c\lambda_0\left(m_0^2-m_{\rm scal}^2\right)}.
}
These expressions indeed give an accurate description of the regime $m_0^2\sim m_{\rm scal}^2$, as shown on \Fig{fig_dressing}. In fact, we see that it remains a good approximation for $m_0^2 < 0.7 \mu_0^2$. Incidentally, it follows from the above analysis that the minimum screening mass is obtained for\footnote{It is interesting to compare the values of $m_{\rm min}^2-m_{\rm scal}^2$ obtained here and in Ref.~\cite{Cyrol:2016tym} (keeping in mind that the respective solutions may, in fact, be of a different nature). Indeed, we expect the quadratic contributions in the setup of that reference to cancel out in this difference. For the set $m_0/\mu_0=0.39$ and $\lambda_0=0.26$, with $\mu_0=1\,{\rm GeV}$, which describes well the SU($3$) lattice data in $d=4$ (see Appendix~\ref{appsec:lattice}), we get $m_{\rm min}^2-m_{\rm scal}^2\approx m_{\rm scal}^2\approx0.05\,{\rm GeV}^2$. This one-loop result falls in the same ballpark as the value $m_{\rm min}^2-m_{\rm scal}^2\approx0.01\,{\rm GeV}^2$ quoted in Ref.~\cite{Cyrol:2016tym}. We expect that the difference between these two values mainly originates from the different definitions of the gluon mass parameter and from logarithmic corrections due to the different scales at which the quantities $m^2_{\rm scal}$ and $m^2_{\rm min}$ are defined ($1\,{\rm GeV}$ here versus $15\,{\rm GeV}$ there).}
\eq{\label{eq:minmass}
m_{\rm min}^2\approx2m_{\rm scal}^2,
}
independent of the constant $c$.

Further away from the separatrix, we observe a linear rise $G^{-1}(0)\sim a + b m_0^2$, with $b\approx 1$. Again, this is the expected behavior at asymptotically large masses, where we find, from Eqs.~\eqn{eq:behavior} or \eqn{eq:FPlT}, $G^{-1}(0)\approx m_0^2$.

That the screening (or infrared) mass decreases for increasing mass parameter $m_0^2$ is not the standard expectation in a massive theory, as pointed out in Ref.~\cite{Cyrol:2016tym} (although we stress again that the status of the control mass parameter used in that work is different from the present one). The authors of this reference have proposed to interpret this as a signature of a ``confining'' phase as opposed to a ``Higgs--like'' phase, where the screening mass increases as a function of the control mass parameter. In particular, this would restrict the range of parameter space where the massive model provides a sensible realization (or gauge-fixed version) of Yang-Mills theory. If the relevant region of parameter space is small enough, this would essentially fix the mass parameter in a unique way, leaving only the coupling as free parameter, just as in the original Yang-Mills theory in the Landau gauge. 

Our findings do not really support this scenario in the present setup. Although we do find two distinct regimes for the gluon screening mass as a function of the mass parameter $m_0^2$, they appear as a simple consequence of the fact that the former diverges when the latter approaches the scaling value. Moreover, there is no sign of a sharp transition between the two regimes. Finally, we find that the range \eqn{eq:minmass} for which the screening mass is a decreasing function of $m_0^2$ is not particularly small. 

\begin{figure}[t!]
  \centering
  \includegraphics[width=.9\linewidth]{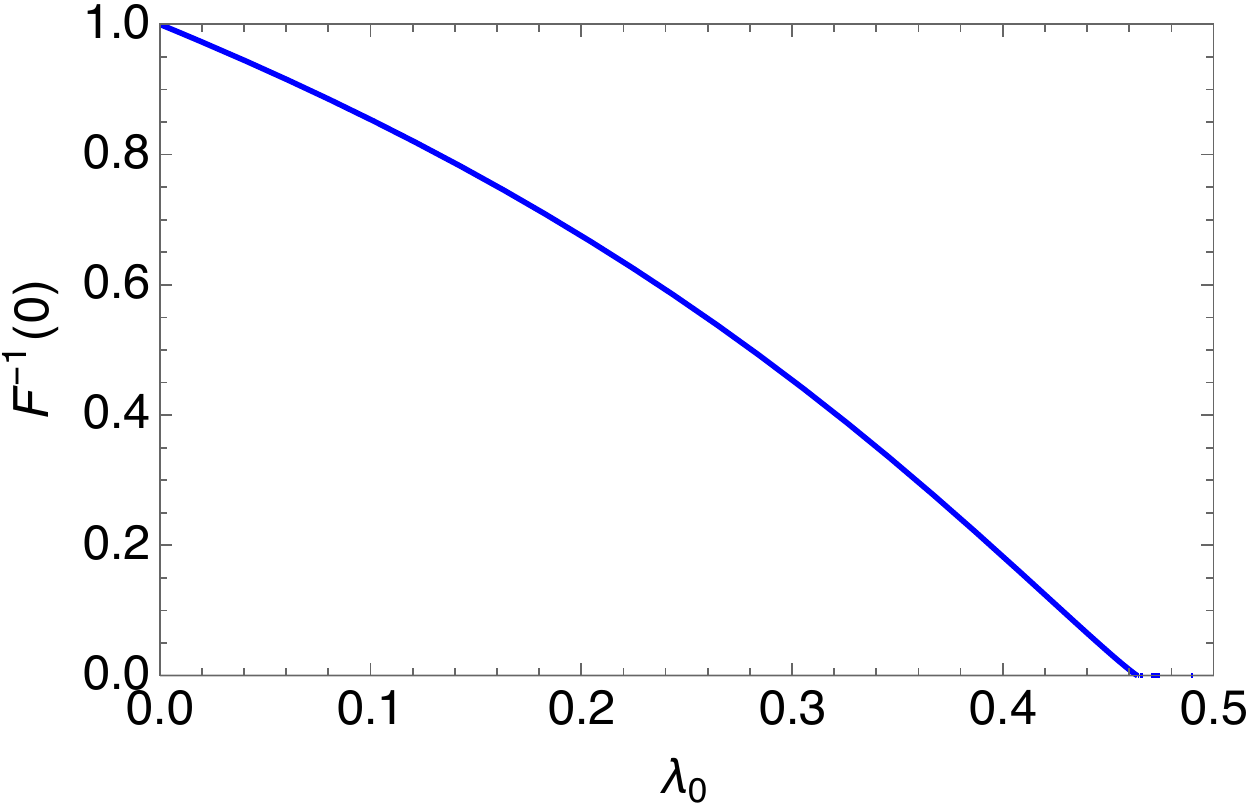}\\
  \caption{The inverse ghost dressing function at vanishing momentum as a function of the coupling $\lambda_0$ for $m_0/\mu_0=0.39$ in $d=4$ (this describes well the SU($3$) lattice data for $\lambda_0=0.26$; see Appendix~\ref{appsec:lattice}). It vanishes at a critical value of the coupling, $\lambda_c\approx0.46$, corresponding to the separatrix (scaling solution).}
  \label{fig_Fvslambda}
\end{figure}

We end this subsection by mentioning that the present analysis sheds some light on the question raised in the literature \cite{Boucaud:2008ji,Fischer08} as to whether the transition from decoupling to scaling solutions is controlled by the inverse ghost dressing function at vanishing momentum $F^{-1}(0)$ or by the value of the coupling $\lambda_0$. It is clear from the RG flow diagram of \Fig{fig_flow} that one can go from a given decoupling solution (any trajectory in the infrared-safe phase) to the scaling solution (the separatrix) by tuning the value of the coupling $\lambda_0$ as long as $\tilde m_0^2<\tilde m_*^2$. In that case, there is indeed a critical coupling for which the scaling solution is reached. We illustrate the one-to-one relation between $F^{-1}(0)$ and $\lambda_0$ at one-loop order in \Fig{fig_Fvslambda}.

\subsection{Spectral positivity violation}

{ Another property of interest is the
spectral positivity violation of the gluon propagator. 
A propagator which satisfies the K\"all\'en-Lehmann representation with a positive spectral function can be shown to be a monotonously decreasing function of (Euclidean) momentum and to have a positive (Euclidean) time Fourier transform \cite{Cucchieri:2004mf}. This allowed both lattice simulations and analytical methods to show that the Landau gauge gluon propagator violates reflexion positivity, in line with the fact that (massive) gluons cannot be asymptotic states. In particular the gluon propagator is clearly not monotonous in two and three dimensions \cite{Cucchieri:2003di,Cucchieri:2004mf,Maas:2007uv,Cucchieri:2008fc}. In four dimensions, lattice data seem to indicate a nonmonotonous behavior too, though in a less conclusive way \cite{Bogolubsky09}.

As first pointed out in Ref.~\cite{Tissier:2010ts},
this feature is correctly captured in the present model
by a genuine perturbative calculation at one-loop order: although the tree-level massive gluon propagator is
a monotonously decreasing function of momentum, the
relevant nonmonotonicity is generated by quantum fluctuations. Here, in order to characterize the positivity violation, we shall consider the momentum at which the gluon propagator has a maximum, to which we shall refer as the scale of positivity violation, following the authors of Ref.~\cite{Cyrol:2016tym}.}

This remains valid with RG improvement, and it is actually a feature of all infrared-safe solutions in the present renormalization scheme at one-loop order. For instance, we have, from \Eqn{eq:glueprop},
\eq{
\frac{d\ln G(p)}{d\ln p}=-\frac{(2-\gamma_A)t+\gamma_c}{t+1},
}
where the gamma functions are evaluated at $\lambda=\lambda(p)$ and $t=p^2/m^2(p)$. { For the decoupling solutions, the relevant infrared limit is governed by the massive ($t\to0$) limit of the gamma functions, see \Eqn{eq:behavior}. We get, in $d=4$,}\footnote{For arbitrary $d$, we have, instead,
$$
\left.\frac{dG(p)}{dp^2}\right|_{p\to0}\sim \frac{G(0)}{m^2(p)}\propto p^{4-d}>0.
$$
This reproduces the known linear rising behavior of $G(p)$ at low momentum in $d=3$ \cite{Maas:2011se,Tissier:2011ey}.}
\eq{\label{eq:nonmono}
\left.\frac{dG(p)}{dp^2}\right|_{p\to0}\sim \frac{\lambda_0}{4m^4_0}\ln\left[\frac{m^2(p)}{p^2}\right]>0.
}
As emphasized in Ref.~\cite{Tissier:2010ts}, this increase of the gluon propagator at low momentum is driven by the loop of massless ghosts. We stress that the behavior \eqn{eq:nonmono} is governed by the infrared limit of the gamma functions, where the running coupling tends to zero, thus justifying the use of the one-loop expressions. Hence, we do not expect higher-order corrections to change this conclusion. 

\begin{figure}[t!]
  \centering
  \includegraphics[width=.9\linewidth]{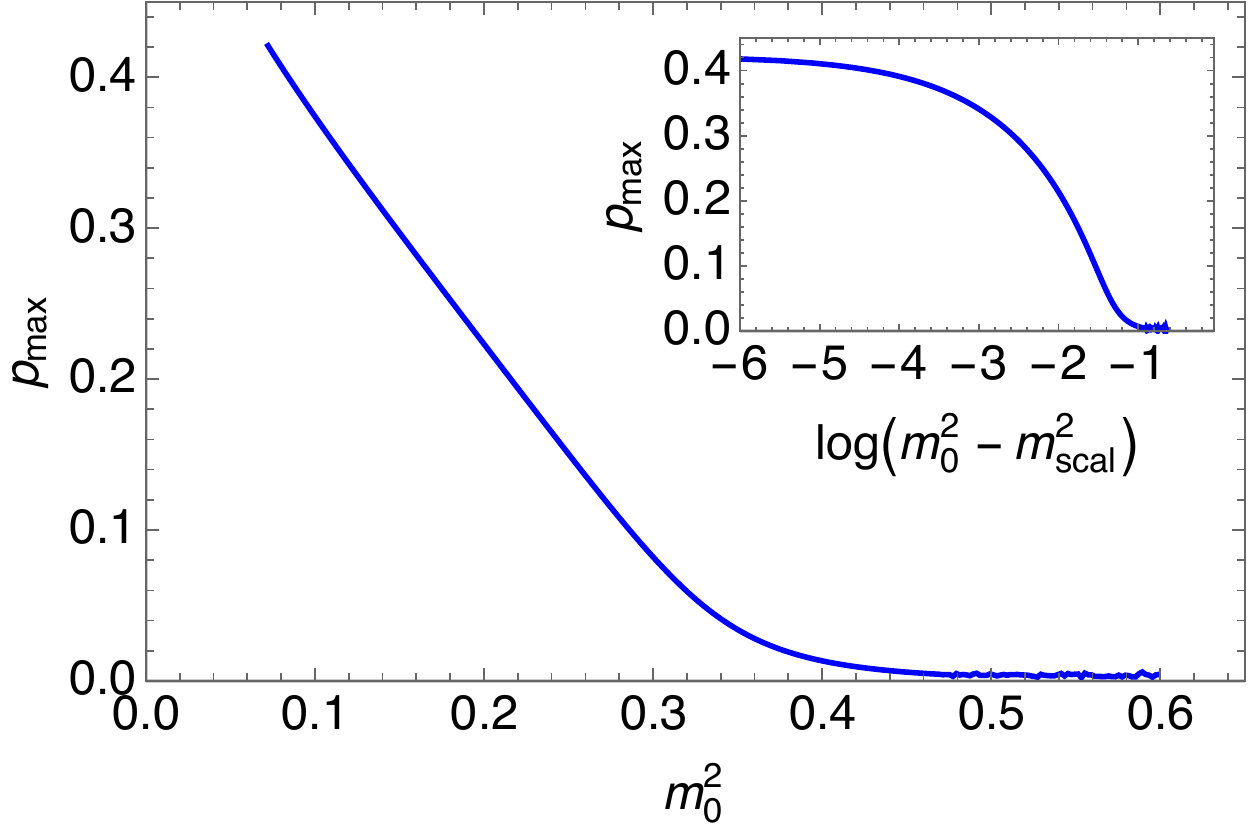}\\
  \caption{$p_{\rm max}$ as a function of $m^2_0$ (units of $\mu_0$) for $\lambda_0=3/\pi^2$. The curve starts on the critical line, i.e. at the value $m_{\rm scal}^2\approx 0.27^2\mu_0^2$ corresponding to the scaling solution. The insert shows the same with a logarithmic horizontal scale.}
  \label{fig_pmax}
\end{figure}

As for the case of the screening mass discussed above, the authors of Ref.~\cite{Cyrol:2016tym} have argued that positivity violation might be present only for a restricted range of parameters, corresponding to the ``confined'' phase, while it would be absent in the ``Higgs'' phase. This is not supported by the present analysis. Based on the above argument, we expect positivity violation to be a common feature of all decoupling solutions and we do not see any sign of qualitatively distinct phases. We do observe though two quantitatively distinct regimes separated by a smooth crossover, where the scale of positivity violation---measured by the position $p_{\rm max}$ of the maximum of $G(p)$---changes from being of order $1$ in the present units\footnote{A typical scale of the problem to compare with is the position of the perturbative Landau pole of the massless FP theory. At one-loop in $d=4$, the latter is $\Lambda_{\rm L}/\mu_0=\exp\{-3/(22\lambda_0)\}\approx 0.64$, with the present parameters.} for $m_0^2$ close to the separatrix, to $p_{\rm max}\ll1$ for larger values of the mass parameter. This is shown in \Fig{fig_pmax}.

We observe a linear decrease of $p_{\rm max}$ as a function of $m_0^2-m_{\rm scal}^2$ from its maximum value at the scaling solution to negligible values, with a transition at about $m_0^2\approx 5 m_{\rm scal}^2$ for the set of parameters used in this figure. This reproduces qualitatively the results of the FRG study of Ref.~\cite{Cyrol:2016tym}. There, the authors mention a power law decrease with an exponent $1.95$. However, this concerns the behavior close to the crossover between the two regimes mentioned above. We have checked that this exponent is not inconsistent with our results in the appropriate region $m_0^2\in[0.3,0.5]$; see the insert in \Fig{fig_pmax}.\\

 To conclude, the model presents an actual phase transition between the Landau-pole phase\footnote{This phase, characterized by $m^2_0<m^2_{\rm scal}$, is referred to as the  ``Coulomb'' phase in Ref.~\cite{Cyrol:2016tym}.}  to the infrared-safe phase, with a boundary characterized by the scaling trajectory. However, we find no sign of qualitatively distinct ``confined'' or ``Higgs'' solutions in the infrared-safe phase, as advocated in \cite{Cyrol:2016tym}, but, rather, a smooth crossover between quantitatively distinct regimes. In this respect, it is interesting to see where the lattice data sit in this picture. As recalled in the Appendix~\ref{appsec:lattice}, the SU($3$) lattice data in $d=4$ are well described in the present approach at one-loop order for the parameters $m_0^2/\mu_0^2=0.39^2=0.1521$ and $\lambda_0=0.26$ at the scale $\mu_0=1\,{\rm GeV}$. For this value of the coupling, the value of the the mass parameter on the separatrix is $m_{\rm scal}^2=0.053\,{\rm GeV}^2$, so that we have $m_0^2>m_{\rm min}^2=0.106\,{\rm GeV}^2$, and the lattice results lie slightly outside the range delimited by \eqn{eq:minmass}. Still, they are well within the regime where the scale of spectral positivity violation is appreciable: We find $p_{\rm max}=0.17\,{\rm GeV}$. Although there is some latitude in describing the lattice data with slightly different parameter sets, this illustrates the fact that there is no clearly separated phase of confining solutions.

\subsection{Longitudinal sector and BRST symmetry}
\label{sec:BRST}

 Finally, we discuss the issue of the possible realization of the BRST symmetry mentioned in the Introduction. The regularization procedure employed in nonperturbative continuum approaches explicitly breaks the standard BRST symmetry of the FP Lagrangian and the possible realization of a BRST symmetric solution requires that their exists a particular value of the parameters---in particular of the gluon mass (counter)term---which exactly cancels the BRST breaking contributions. In principle, establishing whether this is the case or not requires one to analyze the (modified) ST identities in presence of the regulator, which involve, in particular, the longitudinal gluon sector \cite{Ellwanger:1994iz,Lerche:2002ep,Fischer08,Cyrol:2016tym}. This is a difficult task in practical nonperturbative calculations, where one typically focuses on the transverse sector. 

In contrast, the present setup offers an easy access to the longitudinal sector and to the modified ST identities of the massive model. We can thus easily test whether there exists a value of $m_0^2$ for which the standard ST identities are satisfied and, in particular, whether our scaling solution is BRST symmetric. We must stress though that this is a slightly different question from the one above because, in the present case, the standard BRST symmetry is broken only by the tree-level mass term, not by the regulator.

The two-point vertex function (inverse propagator) in the gluon sector admits the following decomposition\footnote{We work here directly in the Landau gauge with the Lagrangian \eqn{eq:model}, where the inversion of the two-point vertex function must be done in the sector $(A,h)$. The $h$ sector being only linearly coupled to the gluon field, it does not receive any loop correction. In particular, it follows that $\Gamma^{(2)}_{hh}(p)=0$, from which one easily checks that the gluon propagator is exactly transverse.}
\eq{
\Gamma_{\mu\nu}^{(2)}(p)=P_{\mu\nu}^\perp(p)\Gamma_\perp(p)+P_{\mu\nu}^\parallel(p)\Gamma_\parallel(p),
}
where $\Gamma_\perp(p)=G^{-1}(p)$. The present model possesses a modified, non-nilpotent BRST symmetry which implies the  identity $\Gamma_{B\parallel}(p)=m_B^2F_B(p)$ for bare quantities  \cite{Curci:1976bt,Tissier:2011ey}. In particular, the standard, nilpotent BRST symmetry of the (massless) FP Lagrangian implies that the gluon vertex function is exactly transverse in the Landau gauge. For renormalized quantities, the above identity becomes $\Gamma_\parallel(p)=m^2F(p)Z_{m^2}Z_AZ_c$. In the present renormalization scheme, this translates into
\eq{\label{eq:longitudinal}
\Gamma_\parallel(p)=m^2_0F(p),
}
which can be equivalently written as
\eq{
\frac{\Gamma_\parallel(p)}{\Gamma_\perp(p)}=m_0^2G(p)F(p)=\frac{\tilde m^2(p)}{1+\tilde m^2(p)}.
}
We see that the modified ST identity of the massive theory implies that the longitudinal gluon self-energy is completely controlled by the ghost dressing function. It follows that, strictly speaking, the only case where the nilpotent BRST symmetry of the FP Lagrangian, namely, $\Gamma_\parallel(p)=0$, is exactly recovered is the massless FP theory $m_0^2=0$. 

Still, there may be situations where the BRST symmetry is approximately recovered even at $m_0\neq0$. This is, for instance, the case in the UV limit, where $\tilde m^2(p)\ll 1$, and the massive theory reduces to the massless one. In this regime, the corresponding ST identity is approximately recovered in the sense that $\Gamma_\parallel(p)/\Gamma_\perp(p)\to0$. However, this is not the case at any infrared but nonzero momentum. For decoupling solutions, where $\tilde m^2(p\to0)\gg1$, we recover $\Gamma_\parallel(0)=\Gamma_\perp(0)$, which is trivial when $\Gamma_{\mu\nu}(p)$ is analytic at $p\to0$, and the BRST symmetry is explicitly broken. For the scaling solution, analyticity is violated and we find, instead,
\eq{
\Gamma_\parallel(p\to0)&\sim\frac{m^4_0}{\lambda_0}\frac{\lambda_*}{\tilde m^2_*} p^{2-d},\\
\Gamma_\perp(p\to0)&\sim\frac{m^4_0}{\lambda_0}\frac{\lambda_*}{\tilde m^2_*}\frac{1+\tilde m^2_*}{\tilde m^2_*}p^{2-d},
}
from which we get a nonzero ratio $\Gamma_\parallel(p\to0)/\Gamma_\perp(p\to0)=\tilde m_*^2/(1+\tilde m_*^2)$, so the nilpotent BRST symmetry is not (even approximately) restored. 

It is worth emphasizing that the above argument is valid at all orders of perturbation theory, relying only on the modified ST identity of the massive model and on the present renormalization scheme. Of course it does not imply that any scaling behavior is incompatible with the BRST symmetry, but it provides an explicit example where a scaling solution is not synonymous of BRST symmetry.\footnote{Another such example is given by the original Gribov-Zwanziger scenario, which does yield a scaling solution in the infrared, albeit explicitly breaking the BRST symmetry by restricting the gauge field configurations to the first Gribov region.}  In fact, we can go a bit further and consider generic approximation schemes which respect the modified ST identities of the massive model. In that case, the argument is less stringent, but still constrains the possible (perturbative) scaling solutions compatible with an approximately restored nilpotent BRST symmetry (at the level of the two-point functions). For a generic scaling solution \eqn{eq:scalingexpdef} with $\alpha_F<0$ (and such that $m_0\neq 0$), we have
\eq{\label{eq:mSTIscaling}
\frac{\Gamma_\parallel(p\to0)}{\Gamma_\perp(p\to0)}\propto p^{2(\alpha_F+\alpha_G-1)},
}
from which we conclude that the (approximate) realization of the ST identity of the FP theory in the infrared requires $\alpha_F+\alpha_G>1$, or $m_0=0$.

We stress again that the above analysis is not directly applicable to the setup of nonperturbative continuum methods because the latter involve a supplementary source of BRST breaking and it might well be that, in that case, there exists a set of parameters which would correspond to $m_0^2=0$ in the above language. This issue is yet to be settled.

\section{Comparison with other approaches}
\label{sec:comparison}

In this section, we compare the results of the above RG analysis to those of nonperturbative continuum approaches. In particular, we have seen that, except near $d=2$, our scaling solution is typically not a weak couplings and might require a nonperturbative treatment. The latter is eased by the observation that either DSE, FRG, or HF equations are only slightly modified by the tree-level gluon mass term,\footnote{ The DSEs of the massive model are strictly identical to those of the FP theory up to a tree-level mass term in the equation for the gluon propagator. In particular, the ghost propagator DSE, from which various constraints concerning possible scaling solutions can be obtained \cite{Boucaud06,Boucaud:2008ji}, is unaffected by the gluon mass term. Moreover, the latter can be absorbed in a (finite) redefinition of the subtraction of the quadratic divergence in the gluon propagator DSE. The same is true in the framework of the HF. Finally, because the functional flow equations of the FRG approach only involve resummed vertices, they are identical for the (massless) FP theory and the massive model. The tree-level mass only appears in the initial conditions for the flow.} and we can mostly rely on the extensive literature on the subject. It is worth emphasizing though that it is only meaningful to compare with results which are compatible with the modified ST identities of the massive model, as is our RG analysis. This is, in particular, the case of early analytical studies of the possible scaling solutions of the DSE \cite{Zwanziger:2001kw,Lerche:2002ep}, for which the dominant infrared behavior is independent of the UV sector. 

Instead, the regularization procedure used in numerical studies in either the DSE, FRG, or HF frameworks explicitly breaks both the BRST symmetry of the FP Lagrangian and the modified BRST (mBRST) symmetry of its massive extension. A meaningful comparison thus requires some care, at least as far as scaling solutions are concerned. This is because of the possibility, mentioned earlier, that scaling solutions in such nonperturbative setups may realize the BRST symmetry of the FP Lagrangian due to an exact cancellation of the BRST breaking effects from the regulator and the gluon mass (counter)term whereas, as discussed at the end of the previous section, our scaling solution is clearly incompatible with the BRST symmetry.

\subsection{Analytical studies}

The possible infrared solutions of the coupled DSE for the ghost and gluon propagators have been intensively discussed in the literature~\cite{Alkofer00,Boucaud:2011ug}. Here, we summarize the results of Refs.~\cite{Zwanziger:2001kw,Lerche:2002ep,Boucaud:2008ji} concerning the necessary conditions for possible scaling solutions of the DSE for the ghost and gluon propagators, where the ghost-gluon vertex is approximated by its tree-level expression.\footnote{The tree-level expression of the ghost-antighost-gluon vertex is unaffected by the gluon mass term. The latter only enters as a tree-level contribution to the gluon DSE. This applies more generally to studies where the dressed vertices are modeled by Ans\"atze constrained by the standard ST identities; see, e.g., \cite{Lerche:2002ep,Fischer08,Boucaud:2008ji}. This is because the modified ST identities of the massive model (in the Landau gauge) differ from those of the FP theory only at the level of the two-point vertex functions (see, e.g., Ref.~\cite{Tissier:2011ey}) and are identical for the three-and-higher-point vertices.} Assuming scaling solutions \eqn{eq:scalingexpdef} for $p\to0$, with $\alpha_F<0$, one can safely neglect the gluon loops contributions in the gluon propagator DSE if $\alpha_G>1$. Note that this also makes the complete tree-level propagator contribution, including the mass term, negligible in the infrared.  If, moreover, $\alpha_F>-1$,
the infrared power law contributions to the relevant loop integrals are convergent while the UV contributions are negligible. In that case, the precise form of the UV regulator is irrelevant and a comparison with the previous RG analysis is meaningful.

\begin{figure}[t!]
  \centering
  \includegraphics[width=.8\linewidth]{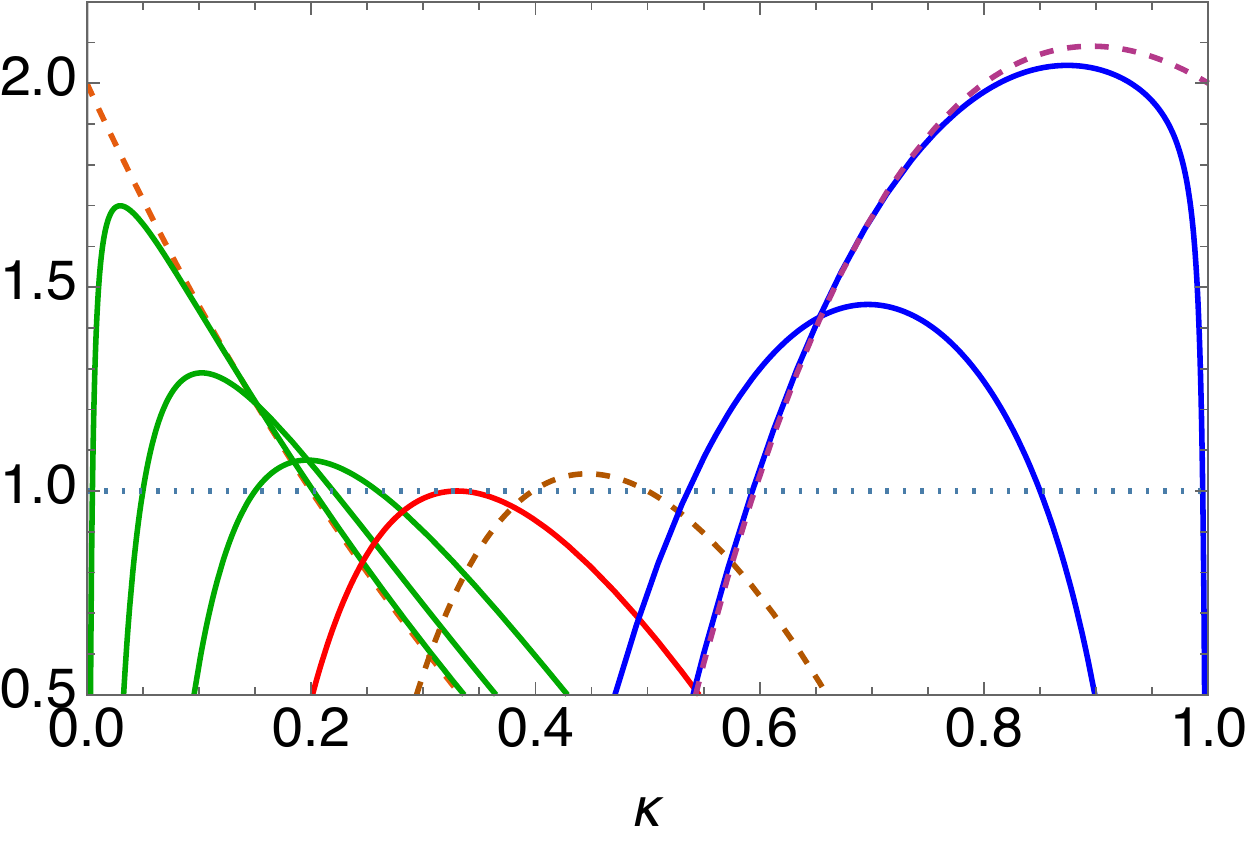}\\
  \caption{Ratio of LHS over RHS of \Eqn{eq:kappaDSE} as a function of $\kappa=-\alpha_F$ for increasing dimensions (from left to right) between $2\le d\le4$. Possible solutions for $\kappa$ are the intersects with $1$. One trivial solution is always $\kappa=(d-2)/2$. Integer dimensions $d=2,3,4$ are shown in dashed lines from left to right. We also show in red the critical dimension $d_c\approx2.66$, for which the two solutions meet and cross. The solution $\kappa=(d-2)/2$ corresponds to the rightmost intersect for $d\ge d_c$ (blue) and to the leftmost one for $d\le d_c$ (green). We see how this solution apparently disappears in $d=2$ and $d=4$ although it always exists for $2<d<4$, as illustrated by the leftmost and rightmost nondashed lines. }
  \label{fig_kappa}
\end{figure}

\begin{figure}[t!]
  \centering
  \includegraphics[width=.85\linewidth]{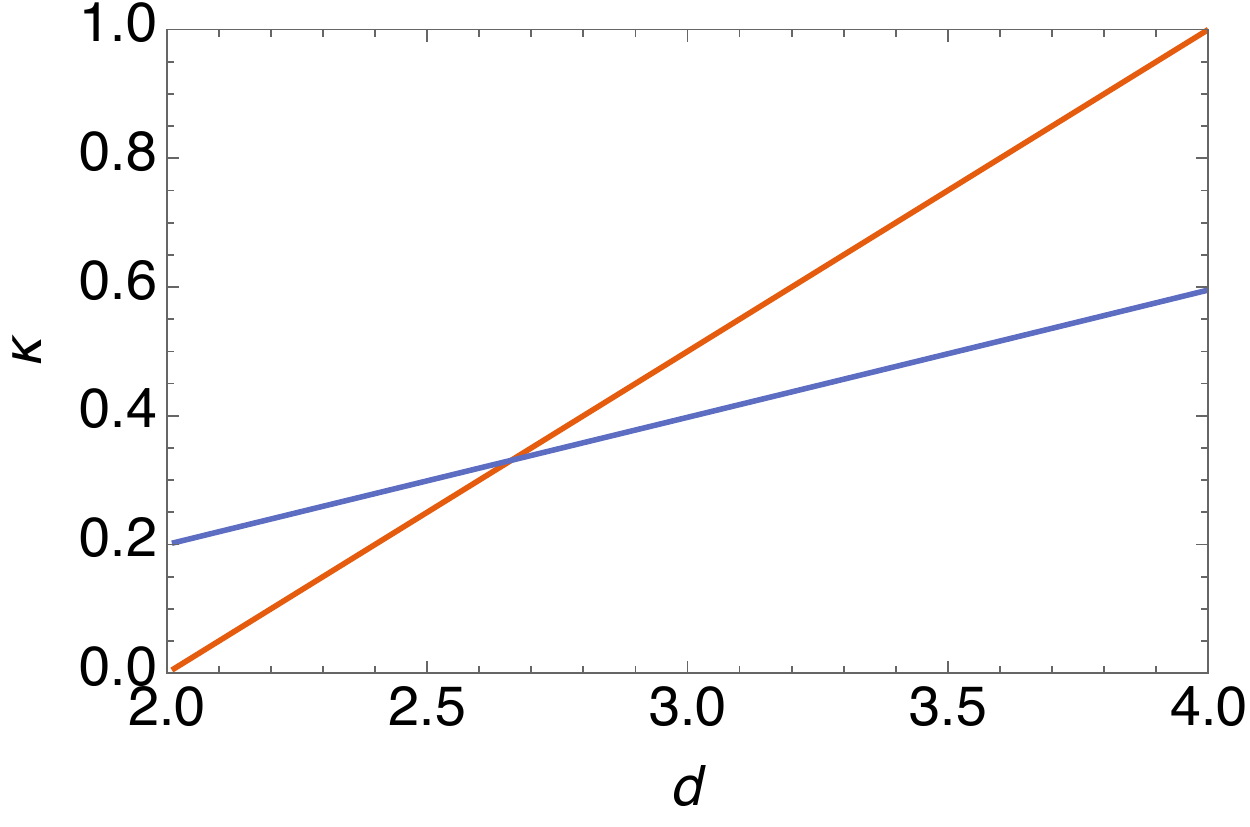}\\
  \includegraphics[width=.85\linewidth]{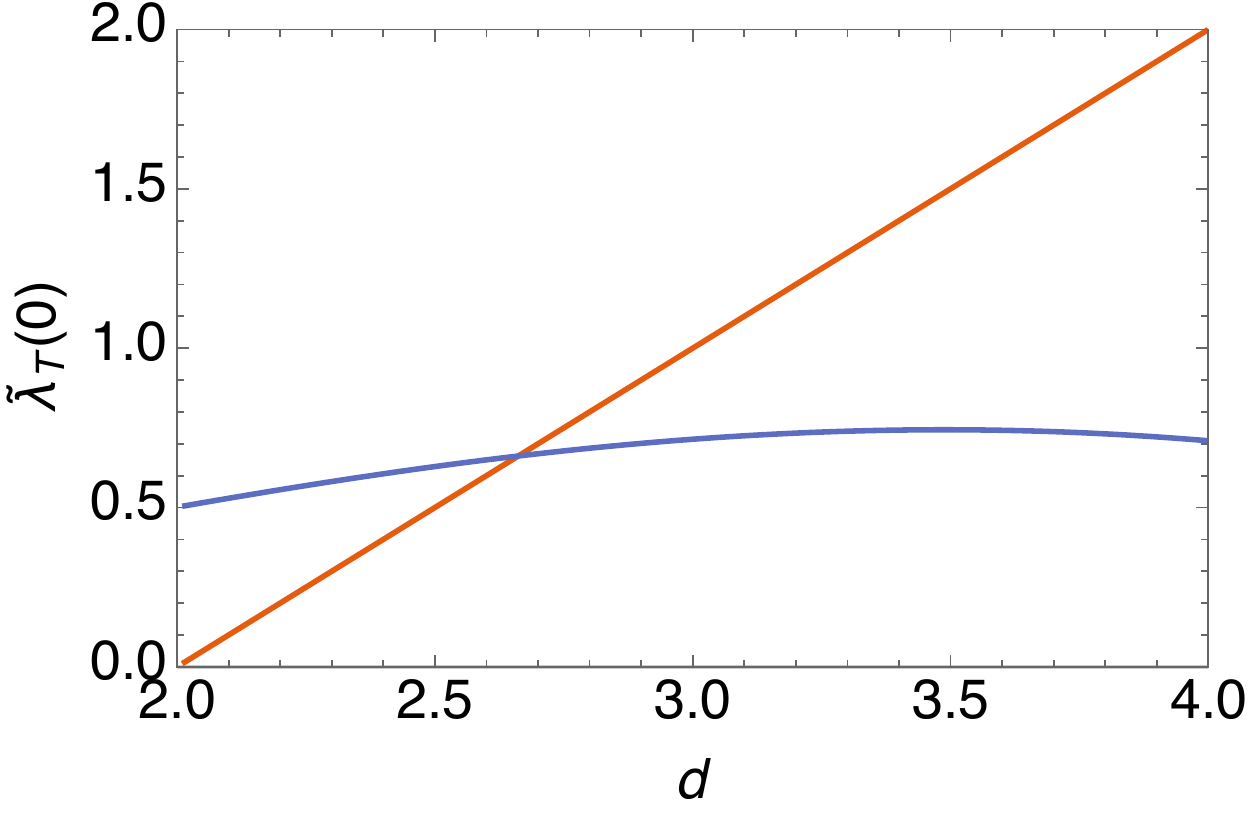}
  \caption{The two possible scaling solutions \eqn{eq:kappastar} and \eqn{eq:kappaother} of the DSEs [top] and the corresponding coupling $\tilde\lambda_T(0)$ [bottom] for dimensions $2\le d\le 4$. The two branches cross at $d=d_c\approx 2.66$.}
  \label{fig_kappa2}
\end{figure}

Under these assumptions, the coupled ghost and gluon DSE yield the following necessary conditions for possible scaling ghost and gluon scaling exponents for $2\le d\le4$:
\eq{\label{eq:DSEscaling}
 \alpha_F=-\kappa\quad {\rm and}\quad \alpha_G=2-\frac{d}{2}+2\kappa,
}
where $\kappa$ lies in the range $(d-2)/4\le\kappa\le d/4$ and solves 
\eq{\label{eq:kappaDSE}
\frac{\Gamma(d-2\kappa)\Gamma(1+2\kappa)}{\Gamma(d/2-\kappa)\Gamma(1+d/2+\kappa)}=\frac{\sin(\pi\kappa)}{(d-1)\sin[\pi(d/2-2\kappa)]}.
}
For $2\le d\le4$, \Eqn{eq:kappaDSE} always has one or two solutions; see \Fig{fig_kappa}. One easily checks that 
\eq{
\label{eq:kappastar}
\kappa_*=\frac{d-2}{2}
} 
is always a solution for $2<d<4$, which, however, spuriously disappears in $d=2$ and $d=4$, where \Eqn{eq:kappaDSE} has singular limits.\footnote{The analytical analysis of Refs.~\cite{Zwanziger:2001kw,Lerche:2002ep}, which yield \Eqn{eq:kappaDSE}, assume that the ghost and gluon DSEs are dominated by the infrared (scaling) regime of the ghost loops and that the UV contributions can be neglected. Possible scaling solutions of this type are restricted to the range $(d-2)/4<\kappa<1$. The branch \eqn{eq:kappastar} does not fall in this category for $d=2$ and $d=4$ and, thus, cannot be excluded by this analysis. In these cases, one should repeat the analysis by taking explicit account of the UV contributions~\cite{Boucaud:2008ji}. Assuming the validity of dimensional regularization, this can be dealt with by taking the limits $d\to2^+$ or $d\to4^-$, in which case, both branches \eqn{eq:kappastar} and \eqn{eq:kappaother} are valid solutions~\cite{Zwanziger:2001kw}. We mention that, in the context of DSE, the solution \eqn{eq:kappastar} has been first obtained in Ref.~\cite{Atkinson:1998zc} in $d=4$.}  The solution \eqn{eq:kappastar} corresponds to the Gribov exponents \eqn{eq:scalingexpd} at the scaling fixed point obtained in the previous RG analysis. The solutions of \Eqn{eq:kappaDSE} are plotted as functions of the dimension $d$ in \Fig{fig_kappa2}. The second solution is well approximated by the linear law \cite{Zwanziger:2003de}
\eq{
\label{eq:kappaother}
\kappa\approx\frac{d-1}{5},
}
which is exact in $d=2$. The two solutions cross each other for a critical dimension $d_c\approx8/3$. In the following, we shall refer to \Eqn{eq:kappastar} as the Gribov branch \cite{Gribov77} and to \Eqn{eq:kappaother} as the von Smekal--Hauck--Alkofer (vSHA) branch \cite{vonSmekal97}.


Because of the scaling relation $2\alpha_G+4\alpha_F=4-d$, the dimensionless coupling $\tilde\lambda_T(p)=p^{d-4}\lambda_T(p)$ [see Eqs.~\eqn{eq:conventionalTaylor} and \eqn{eq:rescaledtilde}] goes to a constant whose value is fixed by the self-consistency of the scaling Ansatz as \cite{Lerche:2002ep,Boucaud:2008ji} 
\eq{\label{eq:DSEcoupling}
\tilde\lambda_T(p=0)=\frac{g^2N}{(4\pi)^{d/2}\Gamma(d/2)}=\frac{1}{\Gamma(d/2) I(d,\kappa)},
}
with the function
\eq{
I(d,\kappa)=\frac{1}{2}\frac{\Gamma^2(d/2-\kappa)\Gamma(1-d/2+2\kappa)}{\Gamma(d-2\kappa)\Gamma^2(1+\kappa)}.
}
To make contact with our previous RG analysis, we evaluate the coupling \eqn{eq:DSEcoupling} for the Gribov branch \eqn{eq:kappastar}. We get
\eq{\label{eq:standardtaylorSDE}
\tilde\lambda_T^*(0)=\frac{1}{\Gamma(d/2) I(d,\kappa_*)}=d-2,
}
to be compared to the one-loop estimate from the preceding section, $\tilde\lambda_T^*(0)\approx 1.06$ in $d=4$ and $\tilde\lambda_T^*(0)\approx 0.95$ in $d=3$. Such a qualitative agreement is remarkable in regard of the strong values of the coupling. Moreover, this agrees exactly with our one-loop result \eqn{eq:lambdatildestar2d} in the perturbative limit $d\to2$. We plot the coupling \eqn{eq:DSEcoupling} for the two solutions of \Eqn{eq:kappaDSE} in \Fig{fig_kappa2}. It is interesting to note that the Gribov branch \eqn{eq:kappastar} corresponds to a strong coupling for dimensions $d\gtrsim3$ whereas the vSHA branch \eqn{eq:kappaother} always yields a moderate coupling $0.5\le\tilde\lambda_T(0)\lesssim0.75$. We mention that these values do not seem to change much with more involved Ans\"atze for the ghost-gluon vertex \cite{Lerche:2002ep,Fischer:2006vf,Fischer08}. The fact that we do not find it in our previous RG analysis at one-loop order suggests that it may be a higher loop effect, that it may require a different renormalization scheme, or that it corresponds to a genuine nonperturbative solution.

Finally, we can easily understand why only the Gribov branch is possible within our perturbative RG approach, at least for $d>d_c$; see \Fig{fig_kappa2}. This is because the other solution \eqn{eq:kappaother} is, in fact, incompatible with Eqs. \eqn{eq:ghostprop} and \eqn{eq:glueprop}. Indeed, the latter imply
\eq{\label{eq:consrgscheme}
m_0^2G(p)F(p)=\frac{\tilde m^2(p)}{1+\tilde m^2(p)}.
}
For $\tilde m^2(p)\ge0$, the combination $G(p)F(p)$ is thus bounded, $0\le m_0^2G(p)F(p)\le 1$. Now, for any given scaling solution, we have $G(p)F(p)\sim p^{2(\alpha_G+\alpha_F-1)}$ and, using \eqn{eq:DSEscaling}, 
\eq{\label{eq:hijklmn}
\alpha_G+\alpha_F-1= \kappa-\kappa_*.
} 
We conclude that our scheme is only compatible with scaling solutions such that $\kappa\ge\kappa_*$. For $d\ge d_c$, this selects the Gribov branch \eqn{eq:kappastar}. For $d<d_c$ instead, both the Gribov and the vSHA branches, Eqs.~\eqn{eq:kappastar} and \eqn{eq:kappaother}, are possible in principle. Note that the latter corresponds to \mbox{$\tilde m^2\propto p^{2(\kappa-\kappa_*)}\to0$}, that is, a {\em massless} infrared limit. As discussed below \Eqn{eq:mSTIscaling}, this also corresponds to the case where the BRST symmetry is approximately restored in the infrared.

\subsection{Numerical studies}

In the usual DSE treatment reviewed above, there is {\it a priori} no reason to exclude one or the other set of scaling exponents. One has to resort to numerical calculations to check whether these are actual solutions of the dynamical equations. As far as scaling solution are concerned, existing studies in $d=4$, see, e.g.,  \cite{Boucaud06,Fischer08,Boucaud:2008ji}, report the vSHA branch \eqn{eq:kappaother}, which is corroborated by numerical calculations in the FRG \cite{Fischer08,Cyrol:2016tym} and HF \cite{Schleifenbaum:2006bq} frameworks. The same is true for $d=2$ \cite{Huber:2007kc,Huber:2012zj}, in which case only a scaling solution seems to exist, in agreement with lattice results \cite{Maas:2007uv,Cucchieri:2011um,Cucchieri:2011ig}. It is important to recall though that the DSE analysis of possible scaling solutions misses the Gribov branch \eqn{eq:kappastar} in $d=2$ and $d=4$. Hence, the numerical DSE studies mentioned here, performed directly in these dimensions, may not really be conclusive concerning this branch. Instead, numerical calculations in $d=3$ \cite{Maas:2004se}, where both the Gribov and the vSHA branches are {\it a priori} possible, indeed find both scaling solutions.\footnote{We mention, though, that the status of the scaling solutions in $d=3$ seems not completely settled yet; see, in particular, the recent study of Ref.~\cite{Huber:2016tvc}, which implements an improved truncation scheme.}

At first sight, the comparison of these numerical studies with the previous RG analysis might appear meaningless because, unlike the latter which uses dimensional regularization, the former generically involve an explicit breaking of the mBRST symmetry from the regularization procedure: the regulator is a source of explicit breaking of the BRST symmetry on top of the gluon mass parameter, which results in an explicit breaking of the mBRST symmetry. As emphasized above in the DSE framework, the tree-level gluon mass contribution can be neglected for scaling solutions and there remains only the question of the BRST breaking contributions from the UV regulator. Fortunately, the latter can be exactly projected out by choosing the so-called Brown-Pennington projection parameter $\zeta=d$ \cite{Brown:1988bm}. In that case, the results are independent of the UV regulator and the comparison with the RG analysis of the previous section should be meaningful. 

Interestingly, the vSHA branch is found to disappear for that particular value $\zeta=d$ in DSE calculations in $d=4$ \cite{Fischer:2002hna} and $d=3$ \cite{Maas:2004se}, although not in $d=2$ \cite{Huber:2012zj}. In the case $d=3$, only the Gribov branch remains and, to the best of our knowledge, one cannot exclude that this may also be the case in $d=4$ in a setup where this branch would be found. This suggests that, indeed, in cases where a comparison can be justified, the DSE results qualitatively agree with those of the previous RG analysis, except for the peculiar case $d=2$. As already mentioned, another scenario is that the vSHA branch is genuinely nonperturbative and not accessible by our approach. In fact, there could even exist two families of decoupling solutions continuously connected to the two scaling solutions, one of which would not be accessible by perturbative means.

\section{Summary and conclusions}
\label{sec:concl}

Because of the practical difficulty of constructing a nonperturbative BRST-invariant regularization scheme, existing continuum approaches to Landau gauge YM correlators rely on deformations of the FP Lagrangian. In most cases, a simple subtraction of quadratic divergences in the gluon self energy is implemented, which amounts to a simple massive extension of the FP Lagrangian, the Landau limit of the CF model. In this context, one assumes that there exists a unique value of the (tree-level) gluon mass parameter which exactly cancels the BRST breaking contributions from the regulator, yielding a BRST symmetric solution.  From another viewpoint, the massive model can be seen as a minimal effective gauge-fixed Lagrangian which takes into account the BRST breaking induced by the Gribov problem. An important question arises in both contexts as to what extent this provides a sensible realization of YM theories. 

In the present article, we have studied perturbatively the parameter space of the massive Lagrangian by means of the infrared-safe RG scheme put forward in Ref.~\cite{Tissier:2011ey}. A one-loop calculation produces the main qualitative features obtained in the literature using a variety of nonperturbative continuum methods, with two classes of either infrared safe or infrared singular solutions, separated by a critical line. Infrared-safe solutions yield a decoupling behavior for the ghost and gluon propagators at infrared momenta, governed by an infrared stable fixed point of the RG flow, similar to a high temperature fixed point \cite{Weber:2011nw}. The scaling solution, instead, is governed by a critical fixed point with an infrared unstable direction and yields a scaling behavior with Gribov exponents. The decoupling fixed point is weakly coupled and thus well-described by perturbation theory. The scaling fixed point is weakly coupled only for dimensions $d\to2$ and strongly coupled otherwise, so that the one-loop analysis is questionable. We have shown though that, for our perturbative solution, the scaling exponents are of the Gribov type at all orders of perturbation theory in the present RG scheme under the sole assumption of a (scaling) fixed point at nonzero finite values of $\tilde m^2_*$ and $\tilde\lambda_*$. We have checked that the latter exists at one-loop order for $2\le d\le4$. In $d=2$ it actually merges with the decoupling fixed point and the whole class of infrared-safe RG trajectories disappears. As already mentioned, Weber \cite{Weber:2011nw} finds similar results in a $d=2+\epsilon$ expansion with a different renormalization scheme. 

We have also analyzed, at one-loop order, the dependence of the ghost dressing function at vanishing momentum, of the gluon screening mass, and of the scale of spectral  positivity violation in the gluon propagator with the parameters of the model. This allows us to discuss in a simple and transparent manner various questions raised in the literature concerning, e.g., the relation between the control parameters of DSE and FRG studies, or the existence of a critical coupling corresponding to the scaling solution. We have also discussed the question of the restoration of the BRST symmetry of the FP Lagrangian. We find that, in the present scheme, the scaling solution does not satisfy, even approximately,  the massless ST identities in the infrared. More generally, we further obtain a constraint on scaling exponents for an approximately restored BRST symmetry in the infrared, based solely on the modified ST identities of the massive model. This constraint is not satisfied by the scaling solutions obtained on the literature for $d>d_c\approx 2.66$. Furthermore, we have analyzed the possibility that different regions of parameter space would describe either a ``confined'' or a ``Higgs-like'' phase, as advocated in Ref.~\cite{Cyrol:2016tym}.  We find no sign of an actual phase transition, but a smooth crossover between strongly quantitative distinct regimes. We stress again that the present results apply to the class of solutions where the mBRST symmetry is manifest and which can be reached perturbatively (possibly at infinite order). However, by no means does this exhausts all possible solutions as there may exist genuine nonperturbative solutions not attainable by perturbative means.

\begin{figure}[t!]
  \centering
  \includegraphics[width=.9\linewidth]{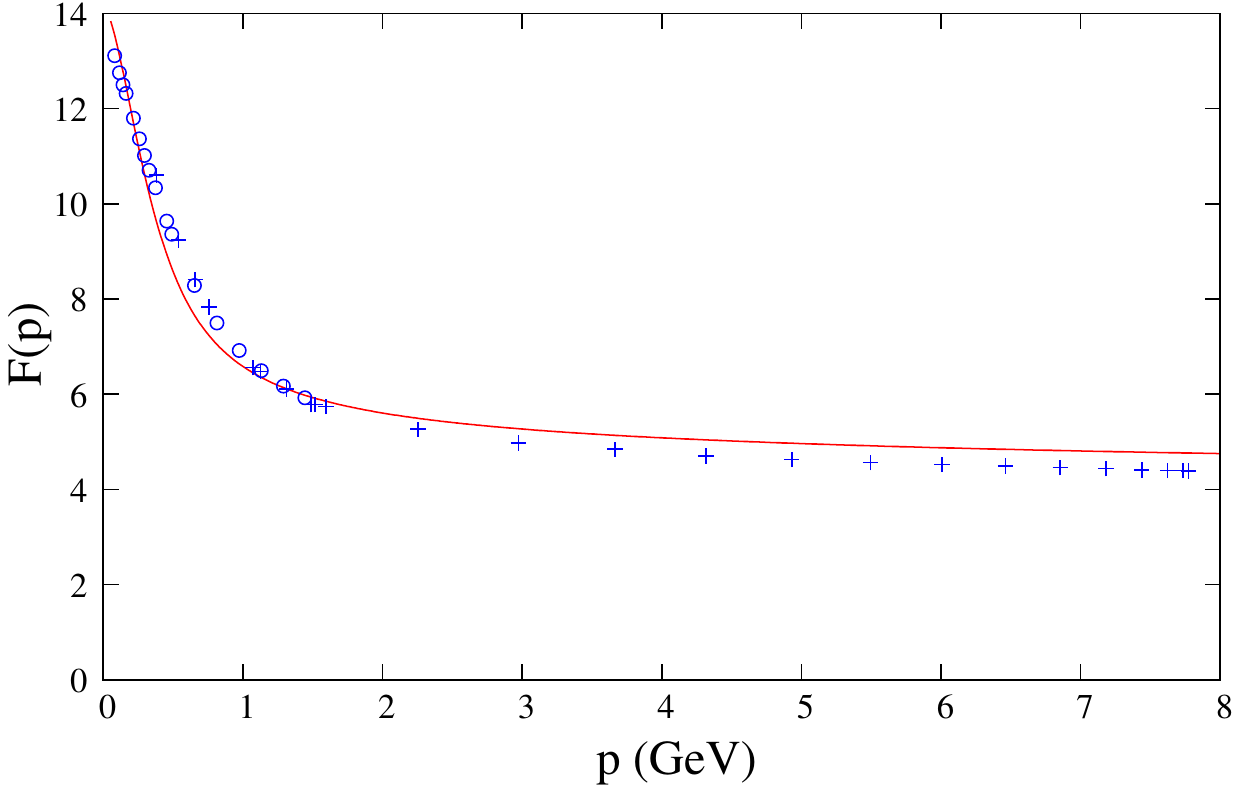}\\
  \includegraphics[width=.9\linewidth]{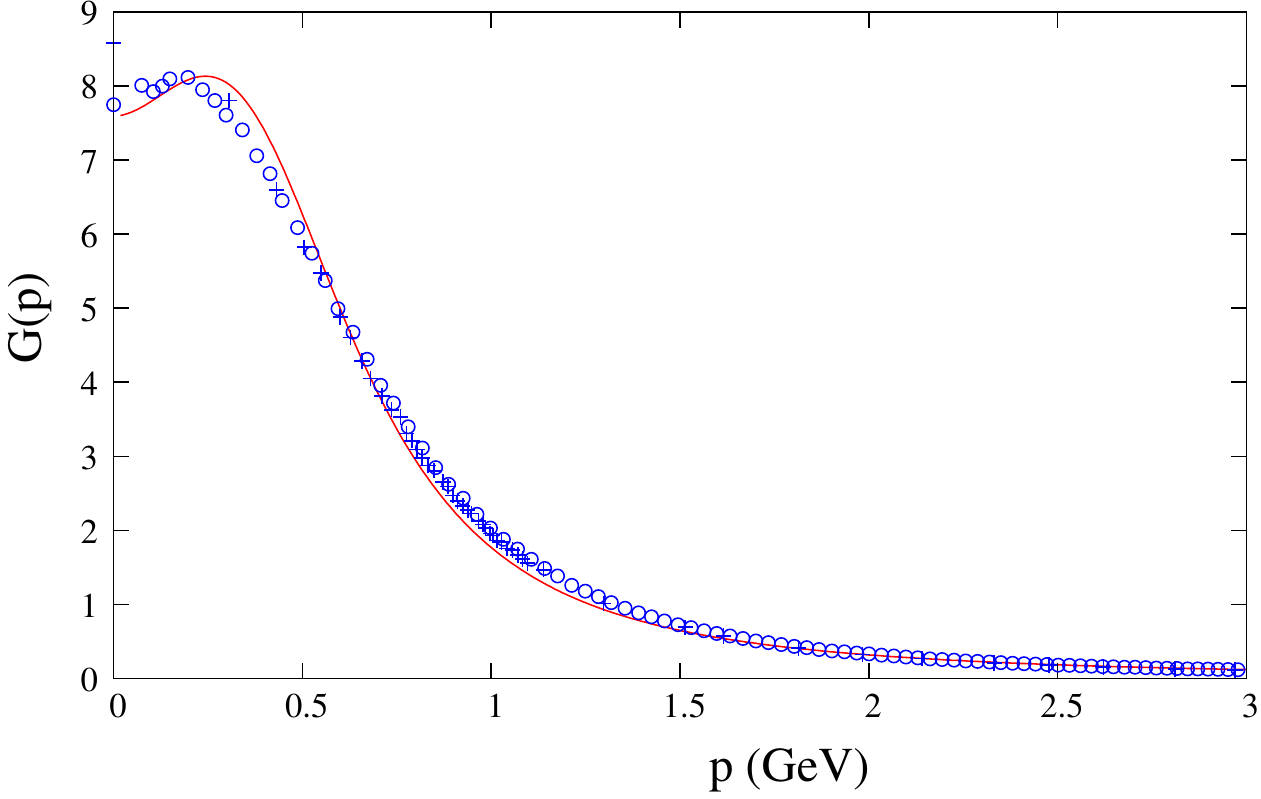}\\
  \caption{The ghost dressing function (top) and the gluon propagator (bottom) for the SU($3$) theory in $d=4$. The lattice data are from Refs.~\cite{Bogolubsky09} (open circles) and \cite{Dudal:2010tf} (crosses), and the curves correspond to the fit from the one-loop infrared-safe scheme. From Ref.~\cite{Tissier:2011ey}.}
  \label{fig_lattice}
\end{figure}

Finally, we have used the existing literature to compare our results for scaling solutions to those of nonperturbative continuum approaches. DSE studies support two scaling solutions in $2< d<4$, described by the exponents \eqn{eq:kappastar} and \eqn{eq:kappaother}. The Gribov branch \eqn{eq:kappastar} corresponds to the scaling fixed point of our RG analysis while the vSHA  branch \eqn{eq:kappaother} is absent, up to one-loop order. Both branches have been found in actual numerical solutions of DSE in $d=3$ \cite{Maas:2004se} and, interestingly, only the Gribov branch is independent of the Brown-Pennington projection parameter. In dimensions $d=4$ and $d=2$, only the vSHA branch is found but we recall that these are somewhat singular cases as far as the Gribov branch is concerned. 

\begin{figure}[t!]
  \centering
  \includegraphics[width=.9\linewidth]{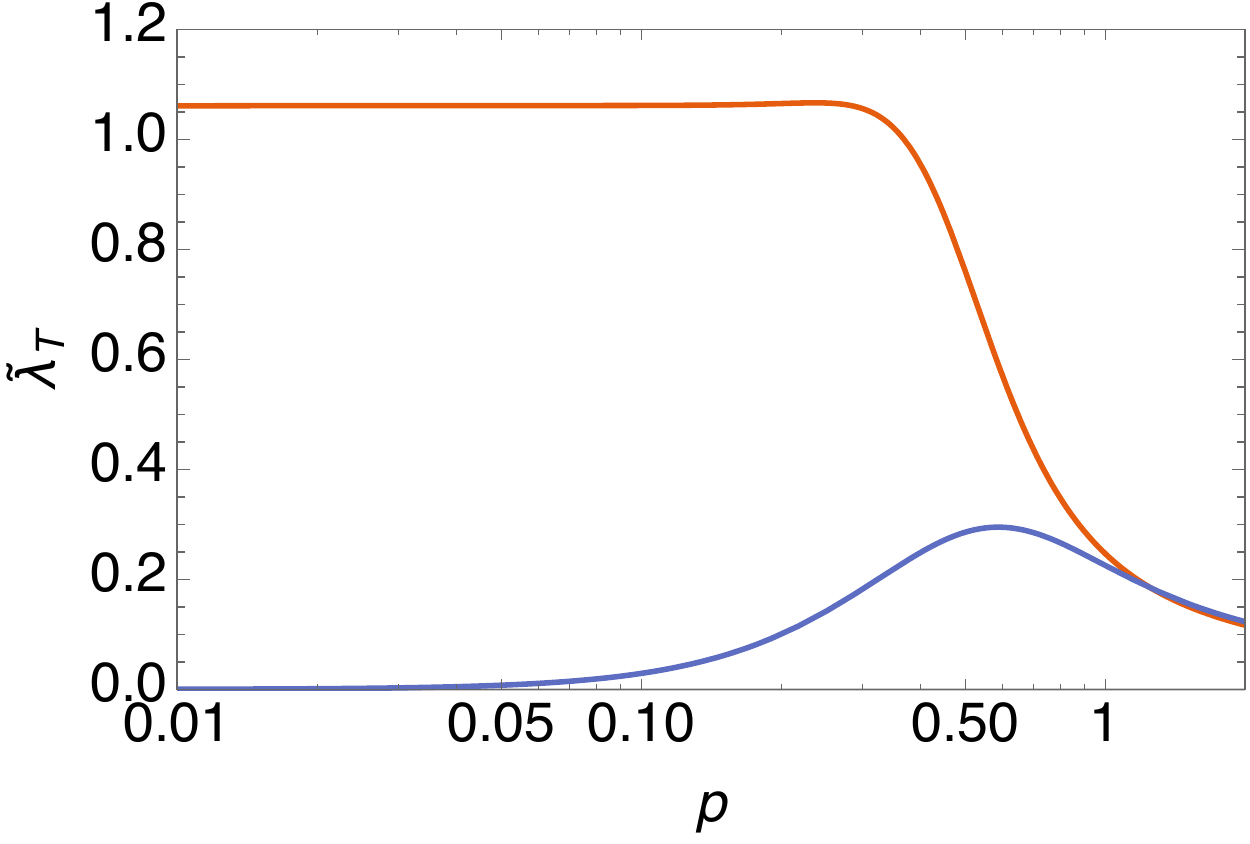}
  \caption{Flow of the rescaled coupling $\tilde\lambda_T$ corresponding to the fit of the lattice data in \Fig{fig_lattice} [$\lambda_0=0.26$ and $m_0=0.39\,{\rm GeV}$, with $\mu_0=1$~GeV], compared to the corresponding scaling solution (red) [$\lambda_0=0.26$ and $m_0=0.23\,{\rm GeV}$].}
  \label{fig_flow-SU3}
\end{figure}

In conclusion, the present RG approach provides a useful tool to investigate the infrared behavior of Landau gauge Yang-Mills propagators, complementary to other continuum approaches. Among the great advantages of this approach are the simplicity of the (perturbative) calculations and the fact that dimensional regularization allows us to control the modified ST identities of the massive model. The RG framework also allows us to select among the possible (decoupling versus scaling) solutions by analyzing the stability of the corresponding fixed points. In this context, the decoupling behavior found in lattice calculations in $d=4$ and $d=3$ is well-described by perturbation theory around the weakly coupled decoupling fixed point \cite{Tissier:2010ts,Tissier:2011ey,Siringo:2015wtx,Machado:2016cij,Weber:2016biv}, as recalled in the Appendix \ref{appsec:lattice}; see Figs.~\ref{fig_lattice} and \ref{fig_flow-SU3}. This decoupling fixed point becomes unstable in $d=2$, where lattice simulations find an infrared scaling behavior with the vSHA exponent \eqn{eq:kappaother}. The fact that, as we have pointed out above, this corresponds to a moderate coupling suggests that it may still be described by (appropriate) perturbative means. However, it remains to be understood how such scaling solution can be obtained in the present RG approach. As we have shown above, this cannot correspond to a fixed point at finite nonzero values of the parameters $\tilde m_*^2$ and $\tilde\lambda_*$ in the RG scheme employed here. Possible ways out---not excluded by the present analysis---could be a nontrivial infrared fixed point at nonzero $\tilde\lambda_*$ but $\tilde m_*^2=0$, or a runaway solutions where $\tilde\lambda\to\infty$ as $\mu\to0$. These are not seen at one-loop order and would require a higher-loop analysis. It could also be that the branch \eqn{eq:kappaother} may not be accessible with the present renormalization scheme and would require either a more appropriate scheme or a more drastic modification of the FP Lagrangian to start with, maybe in the line of the Lifshitz point described in Ref.~\cite{Weber:2011nw}. Finally, as already mentioned repeatedly, we cannot exclude that the scaling behavior in $d=2$ could also be a genuine nonperturbative aspect of the (massive) theory. 

It would also be interesting to clarify the status of the two branches of scaling solutions \eqn{eq:kappastar} and \eqn{eq:kappaother} on the side of DSE/FRG/HF calculations. For instance, as already mentioned, there could be two separate branches of decoupling solutions, each ending on one of the scaling solutions. Finally, it would be of great interest to generalize the existing numerical studies for arbitrary dimension, e.g., along the lines of Ref.~\cite{Dall'Olio:2012zw}, and to investigate the possibility of implementing dimensional regularization, e.g., following Refs.~\cite{Schreiber:1998ht,Gusynin:1998se,Phillips:1999bf}. The transition from a (perturbative) decoupling behavior in $d=4$ to a (possibly nonperturbative) scaling behavior in $d=2$ remains one of the important open questions in the field.

\section*{Acknowledgements}

We acknowledge interesting discussions with R. Alkofer, A. Cyrol, A. Maas, and M. Mitter. We are grateful to J. M. Pawlowski for useful discussions and suggestions.

\appendix

\section{Comparison to lattice results}
\label{appsec:lattice}

Here we simply recall, for completeness,  how the results of the present approach in the infrared-safe renormalization scheme at one-loop order compare with the lattice data \cite{Tissier:2011ey}. The ghost dressing function and the gluon propagator of the SU($3$) theory in $d=4$ dimensions are shown in \Fig{fig_lattice}. The RG improved one-loop results give a good description of the lattice results of Refs.~\cite{Bogolubsky09,Dudal:2010tf} over a wide range of momenta, from the ultraviolet to the (deep) infrared for the set of parameters $\lambda_0=0.26$ and $m_0/\mu_0=0.39$ at the scale $\mu_0=1$~GeV, where the correlators are normalized to the lattice ones so that we have the same definition of GeV.

We also show the corresponding flow of the rescaled coupling $\tilde\lambda_T$, the appropriate loop-expansion parameter, in \Fig{fig_flow-SU3}. We see that it remains moderate throughout the whole momentum range, which justifies the one-loop approximation. We also show the flow for the scaling solution corresponding to the same coupling $\lambda_0=0.26$, which occurs for the critical mass $\tilde m_{\rm scal}^2\approx 0.05\,{\rm GeV}^2$. The trajectory corresponding to these lattice data is also shown (orange curve) in \Fig{fig_flow}.

\section{An instructive toy example}
\label{app:toy}

\begin{figure}[t!]
  \centering
  \includegraphics[width=.9\linewidth]{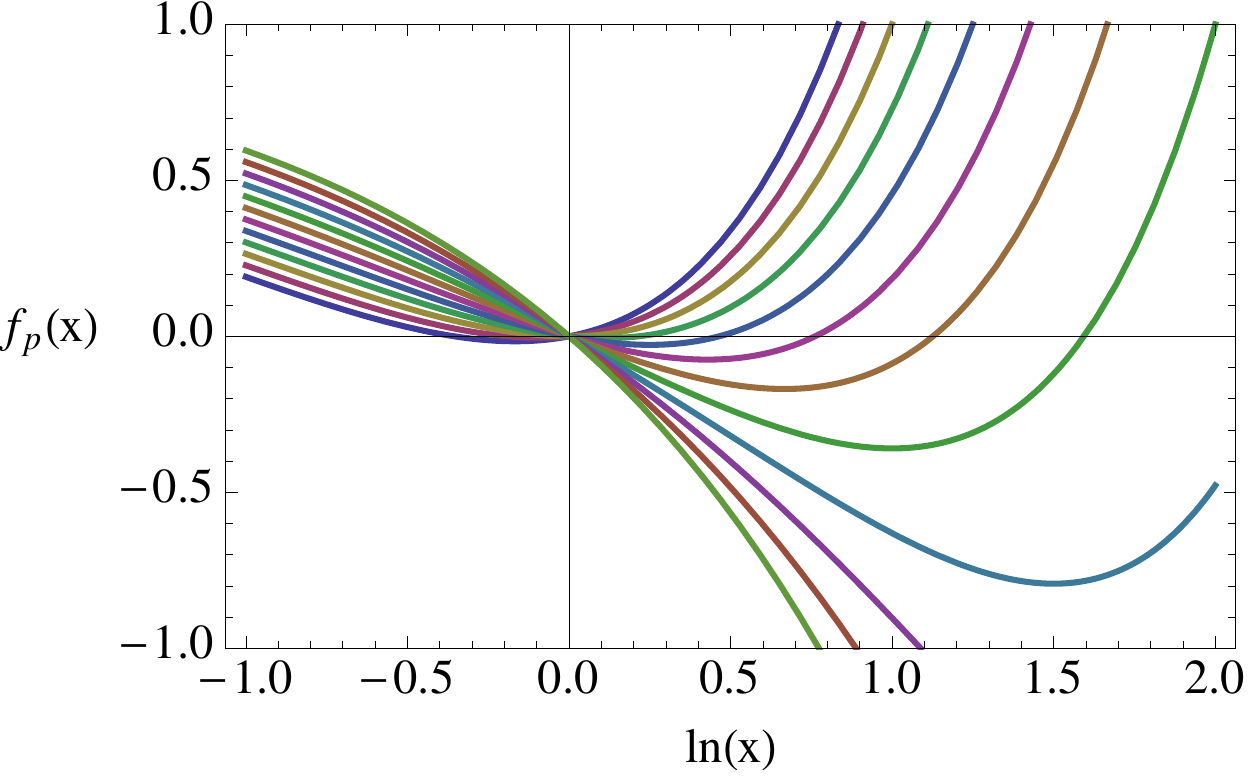}
  \caption{The function $f_p(x)$ as a function of $\ln x$ for $m_B=1$, $a(p)=1$ and decreasing values of $g_B$ from top to bottom on the right of the vertical line. We observe that, for each value of (small enough) $g_B$, there are two solutions, one of which is not expandable in powers of $g_B$.}
  \label{fig:toy}
\end{figure}

In this section, we would like to illustrate on a toy DSE that, for a given theory (that is given values of the bare parameters), they may exist various solutions obeying different renormalization conditions, and that some these solutions may not be accessible through a perturbative expansion. Consider the following equation (that would correspond to a DSE in our toy model):
\beq\label{eq:toy}
x(p) = m_B^2+g_B\,a(p)\,x(p)\,\ln x(p)\,,
\eeq
for a given function $0<a(p)<1$ and with fixed $m_B$ and $g_B>0$ specifying the theory. Writing the equation in the form $0=f_p(x(p))$, we have
\beq
f'_p(x)=-1+g_B a(p)(1+\ln x)\,,
\eeq
which changes from negative to positive at the point $\smash{x=x_0\equiv\exp\{1/(g_Ba(p))-1\}}$. It follows that, as one increases $x$ from $0$ to $\infty$, $f_p(x)$ decreases from
$m_B^2>0$ to $f_p(x_0)=m_B^2-g_B a(p)\exp\{1/(g_Ba(p))-1\}$ and then increases to $+\infty$. Thus for $g_B$ small enough, $g_B a(p)$ is small enough for all $p$, and (\ref{eq:toy}) admits two solutions, which obviously obey different renormalization conditions (for the same $m_B$ and $g_B$). Moreover one of the solutions is always larger than $x_0$ and thus does not admit an expansion in powers of $g_B$, since $x_0\to\infty$ as $g_B\to 0$. In Fig.~\ref{fig:toy}, we show the function $f_p(x)$ for $a(p)=1$, for fixed $m_B$ and decreasing values of $g_B$.

\end{document}